\documentclass[useAMS,usenatbib,dcolumn,usegraphicx,twocolumn,10pt]{mnarxiv}
\usepackage[dvips]{graphicx}
\usepackage{txfonts}
\usepackage{amssymb}
\usepackage[scriptsize,bf]{caption}
\usepackage{float}
\usepackage{placeins}
\usepackage{xspace}
\usepackage{natbib}
\usepackage{hyperref}
\usepackage{textgreek}
\usepackage{multirow}
\usepackage{rotating}
\usepackage[usenames,dvipsnames]{color}
\usepackage[normalem]{ulem}

\newcommand{\A}{u_a}
\newcommand{\B}{u_b}
\newcommand{\km}{\mathrm{km}}
\newcommand{\kms}{\,\km/\mathrm{s}}
\newcommand{\kpc}{\,\mathrm{kpc}}
\newcommand{\msun}{\mathrm{M}_{\odot}}
\newcommand{\pdf}{\textsl{PDF}\xspace}
\newcommand{\rvd}{\textsl{RVD}\xspace}
\renewcommand{\sec}{\mathrm{s}}
\newcommand{\mw}{\textsc{MW}\xspace}
\newcommand{\leoI}{\textsc{Leo I}\xspace}
\newcommand{\fromtokpc}[2]{#1-#2\kpc}

\newcommand{\lcdm}{$\Lambda$CDM\xspace}
\newcommand{\mvir}{M_{\mathrm{vir}}}
\newcommand{\tX}{\textsf{X}\xspace}
\newcommand{\tA}{\textsf{A}\xspace}
\newcommand{\tB}{\textsf{B}\xspace}
\newcommand{\tC}{\textsf{C}\xspace}
\newcommand{\tD}{\textsf{D}\xspace}
\newcommand{\tII}{\textsf{SI}\xspace}
\newcommand{\tI}{\textsf{SII}\xspace}

\newcommand{\avg}[1]{\langle{#1}\rangle}
\newcommand{\br}[1]{\left({#1}\right)}
\newcommand{\disp}[1]{\langle{#1}\rangle}
\newcommand{\e}{\epsilon}
\newcommand{\p}{\mathcal{P}}
\newcommand{\q}{\mathcal{Q}}
\newcommand{\ud}[1]{\mathrm{d}{#1}}

\renewcommand{\sq}[1]{\left[{#1}\right]}
\newcommand{\figref}[1]{Fig.\,\ref{#1}}
\newcommand{\tabref}[1]{{\color{black}Table}\,\ref{#1}}
\newcommand{\eqref}[1]{(\ref{#1})}
\newcommand{\secref}[1]{{\color{black}Sect.}\,\ref{#1}}
\newcommand{\secrefs}[2]{{\color{black}Sects.}\,\ref{#1} and \ref{#2}}
\newcommand{\xxout}[1]{}

\title[]{A
lower bound on the Milky Way mass from general phase-space
distribution function models}
\author[L. Bratek]{
{{\L}ukasz Bratek$^{1}$}, {Szymon Sikora$^{2}$}, {Joanna
Ja{\l}ocha$^{1}$}, {Marek Kutschera$^{3}$}
\\
$^{1}$Institute of Nuclear Physics, Polish Academy of Sciences,
Radzikowskego 152, PL-31342 Krak\'{o}w, Poland\\
$^{2}$Astronomical Observatory, Jagiellonian University, Orla 171,
PL-30244 Krak\'{o}w, Poland\\
$^{3}$Institute of Physics, Jagiellonian University,  Reymonta 4,
PL-30059 Krak{\'o}w, Poland }

\begin{document}
\date{v(1): 8 Aug 2011; \dots; v(this): 5 Sep 2013; Accepted: A\&A 4 Dec 2013}
\pagerange{\pageref{firstpage}--\pageref{lastpage}} \pubyear{2014}

\maketitle

\begin{abstract}
We model the phase-space
      distribution of the kinematic tracers using general, smooth
      distribution functions to derive a conservative lower bound on
      the total mass within $\approx\fromtokpc{150}{200}$.
      By approximating the potential as Keplerian, the
      phase-space distribution can be simplified to that of a smooth
      distribution of energies and eccentricities. Our approach naturally
      allows for calculating moments of the distribution function, such as the radial profile
      of the orbital anisotropy.\\ \indent
      We systematically construct a family of phase-space
functions with the resulting radial velocity dispersion
overlapping with the one obtained using data on radial motions of
distant kinematic tracers, while making no assumptions about the
density of the tracers and the velocity anisotropy
parameter $\beta$ regarded as a function of the radial variable.\\
\indent While there is no apparent upper bound for the Milky Way
mass, at least as long as only the radial motions are concerned,
we find a sharp lower bound for the mass that is small. In
particular, a mass value of $2.4\times10^{11}\msun$, obtained in
the past for lower and intermediate radii, is still consistent
with the dispersion profile at larger radii. Compared with much
greater mass values in the literature, this result shows that
determining the Milky Way mass is strongly model dependent. We
expect a similar reduction of mass estimates in models assuming
more realistic mass profiles.
\medskip
\hrule
\flushleft \textbf{The definitive version \texttt{ A\&A 562, A134 (2014) } is available at\\
\url{http://dx.doi.org/10.1051/0004-6361/201322617}}
\medskip
\hrule
\end{abstract}


\section{Introduction}\label{sec:intro}


The asymptotic value of galactic mass function (total mass) can be
ascertained through studying the radial motions of distant
kinematic tracers, which are regarded as test bodies moving under
the influence of the galactic gravitational field. A primary
quantity for describing a collection of these bodies is a
Phase-space Distribution Function (\pdf) which must be
non-negative everywhere. The function gives rise to various
theoretical secondary quantities such as the radial velocity
dispersion (\rvd), the flattening of the velocity ellipsoid
$\beta$ (the anisotropy parameter), the number density of tracers,
mean velocities, etc, which in general may be functions of space
variables defined as appropriate integrals involving the \pdf. On
comparison with the corresponding quantities from measurements,
the total mass can be inferred.

The accustomed approach to this technique of determining the total
mass is based on Jeans modeling of a stationary collision-less
system of identical bodies in a steady-state
equilibrium\footnote{$\partial_t f\equiv0$ (steady-state);
$d_tf\equiv0$ (equilibrium) for a \pdf $f$}
\citep{1915MNRAS..76...70J}. Jeans concludes his work by saying
that the Galaxy has not yet reached a final steady state on
account of the fact that such a state seems to be inconsistent
with the observed streaming motions. The principal assumption of
this technique requires that the kinematic tracers are relaxed in
the gravitational potential and can be described in terms of a
smooth \pdf.
 However, the
assumption of collision-less equilibrium may not necessarily be
appropriate. In particular, this may be the case both for fast
moving tracers (e.g., propelled by the three-body ejecting
mechanism, which is collisional in its nature) and for most
distant tracers, on orbits characterized by large time scales,
which are likely to be influenced by the material distributed on
larger scales in the Local Group. It is also doubtful that the
very notion of a smooth distribution is applicable to an extremely
rarefied collection of the outermost satellites of the Milky Way
(\mw). Although Jean's modeling is a powerful tool that was
fruitful in many cases, it should be remembered that it has a
limited application.

In recent years, there has been growing interest in the most general
  distributions functions that can be fit to dynamical data with most general and assumption-free constraints on the
  gravitational potential and mass profile. Under the
  assumption that a galaxy is in a steady state,
  \citet{2013arXiv1303.6099M} proposes a new framework of
  inferring the gravitational potential from a discrete realization of
  the
  unknown distribution function provided by the snapshots of a galaxy's
  stellar kinematics.
\citet{2010ApJ...711.1157B} used instantaneous kinematic
{snapshots} of radial distances and velocities to infer the force
{law with} a fully Bayesian inference technique. As an example,
{when} applying it to the solar system, the correct exponent in
the power form of the force law was almost precisely
reconstructed, largely independent of the \pdf of two variables
(energy and eccentricity) with no need to make strong assumptions
about the anisotropy. As noted in {\citet{2010ApJ...711.1157B}},
generalizations of the methods used therein would permit inference
of \mw dynamics  from upcoming surveys, such as the Gaia mission
\citep{2001A&A...369..339P}, or of the mass of the central black
hole.

In this paper, the form of \pdf is also general. Its form is not
assumed except that it should be a function of integrals of motion
to automatically satisfy the Boltzmann equation. Our aim is to
find, in a given gravitational potential that allows for such
integrals, a variety of \pdf{}s consistent with the radial
distance-velocity data. In this preliminary work, we do not make
any assumptions about the form of other secondary observables
derivable from a given \pdf (such as the transverse velocity
dispersion or the profile of the anisotropy parameter) that could
be used to constrain the variety of consistent \pdf{}s. We focus
instead on the resulting estimate of the \mw's minimum total mass.

\medskip

The secondary quantities referred to above are not all
independent. They are constrained to satisfy (moment) Jeans
equations. However, satisfying them is not sufficient {for the
positivity} of the \pdf. In principle, having found a solution,
the positivity condition should be checked separately.
Jeans equations are usually {underdetermined} -- there is
arbitrariness in choosing solutions.  Some of the secondary
quantities must be {assumed} (along with the mass function of the
host gravitational field). This makes the problem of determining
the total mass {model-dependent and} biased by inescapable
degeneracies. The existing mass estimates can differ largely
between each other. The differences lie both on the side of
choosing sample tracers and making model assumptions. In the
context of this under-determinacy, one cannot exclude that masses
predicted by various models can be overestimated. The mechanism of
this overestimation can be understood as follows. The total mass
can be regarded as a functional {defined on} the space of
solutions to the Jeans equation.\footnote{For example, the
gravitational energy of a virilized self-gravitating dust is equal
to minus twice the total kinetic energy. On dimensional grounds,
the binding energy is a product of the total mass squared and the
inverse of some characteristic size determined by the spatial
distribution  of dust. {The} total mass of a virialized system {is
thus a} functional on the phase space.} As said above, various
quantities entering that equation must involve additional
assumptions. These assumptions are introduced in the form of
constraints, {such as} a constant value for $\beta$, a power law
for the number {density, and the} particular form of the
gravitating mass function of the host {potential.} These
constraints impose some indirect restrictions on the \pdf
function. {As} for any functional with {constraints, so also} for
the total mass {functional,} one can expect that the lower bound
may be increased compared to that without constraints.\footnote{A
good example of determining lower bounds with constraints is
provided by the minimum-maximum method of finding consecutive
eigenvalues and eigenvectors of a quadratic form in a finite
number of dimensions or, in an infinitely dimensional function
space, {of} a self-adjoint differential equation
\citep{1953mmp..book.....C}.}

From the above consideration, a natural question arises for
\textit{the lower bound of the \mw mass}. This question motivates
our paper. Our aim is to {show that} with a given set of radial
motions of kinematic tracers, the estimated mass can be much lower
than usually obtained in the literature. We postpone answering
this question until \secrefs{sec:ensemble}{sec:results}.

Another, but related, question concerns the upper bound. It may be
answered more straightforwardly. {We} consider the point mass
potential -- the asymptotics of any physical mass distribution.
The time average $\disp{rv_r^2}_t=\frac{1}{2}GMe^2$ for a single
elliptical orbit with eccentricity $e$, suggests that
$\disp{v_r^2}\sim r^{-1}$ for almost circular orbits. Assuming
this behavior of $\disp{v_r^2}$ in the spherical Jeans equation
$\frac{r\rho_{,r}}{\rho}+2\beta=
-\frac{r}{\disp{v_r^2}}\br{\Phi+\disp{v_r^2}}_{,r}${,} along with
the vanishing mean velocities, constant
$\beta=1-\frac{\disp{v_{\theta}^2}+
\disp{v_{\phi}^2}}{2\disp{v_r^2}}${,} and
$\rho\sim\rho_o\br{r_o/r}^{3+\varepsilon}$ ($\varepsilon>0$), we
conclude that asymptotically, when $r\,\Phi\sim{-G M}$, the mass
estimate for almost circular orbits is{}
\begin{equation}\label{eq:Simple_Estimator}
M\sim\mu\,G^{-1}\,r\disp{v_r^2}\quad \mathrm{with}\quad
\mu=4+\varepsilon-2\beta>4-2\beta>2.\end{equation} This result
holds without the need for the phenomenological explicit
assumption that $\disp{v_r^2}\sim r^{-1}$. A general integral of
the spherical Jeans equation found in \citet{2009ApJ...701.1500A}
under similar assumptions about $\beta$ and $\rho${,} reduces in
the point mass potential to $\disp{v_r^2}=\frac{G
M}{4-2\beta+\varepsilon}r^{-1}+C\, r^{4-2\beta+\varepsilon}$ (in
our notational convention), where $C$ is an integration constant.
In the asymptotical regime we work in, the appropriate boundary
condition is to require that $\disp{v_r^2}$ should be finite in
the limit $r\to\infty$, which for $4-2\beta+\varepsilon>0$ implies
that $C=0$, again giving \eqref{eq:Simple_Estimator}. {Result}
\eqref{eq:Simple_Estimator} says that an almost flat
$r\disp{v_r^2}$ profile could be explained by arbitrary large $M$
($\beta$ and $\varepsilon$ are not controlled by measurements at
large radii\footnote{It is not inconceivable that we can make in
the nearest future more precise measurements of parameters $\beta$
and $\varepsilon$ at the outermost radii, in which case solutions
of the Jeans equation would be more robust.}). {T}here is no
apparent upper bound for the mass. This will become even more
evident in \secref{sec:results}.

In the asymptotic consideration above, the presence of arbitrary
parameters (such as $\varepsilon$ and $\beta$) represents the
freedom in choosing solutions in more general situations. As we
have seen, the relation between the mass function and the \rvd
profile must be arbitrary to some extent, because the kinematics
is not entirely linked to the mass distribution. This is the
important degeneracy of the Jeans problem:  the inferred mass
depends on the assumed model parameters (especially $\beta$ in the
above example). In other words, the asymptotics of solutions is
not entirely fixed by measurements in the interior, so some
analytic continuation of solutions must be assumed.  This
indeterminacy is physically clear. First, the motions are not
entirely due to the massive (monopole) term in the potential.
Higher multipoles also affect the motion, despite being massless
components of the gravitational field. Second, in a given
potential one can consider infinitely many ensembles of test
bodies with various stationary velocity dispersion profiles
(evolved from various sets of initial data). There are some
conserved quantities (e.g., total energy) that constrain the
evolution in the phase space, therefore, the (collisionless)
relaxation to a stationary state cannot lead to a universal
dispersion profile. The profile should be a functional of the
initial data.

\medskip

With general solutions to the Jeans problem, the estimated mass
value can be reduced effectively. We show this in
\secref{sec:ensemble} in the approximation of a point mass field.
The idea of studying the motions of external tracers in a
point-mass approximation should not be surprising. It was already
considered over 30 years ago by \citet{1981ApJ...244..805B} and
applied to several external galaxies. The authors proposed an
estimator for the galactic mass of the form $\frac{C}{G}\avg{v_z^2
R}$, where the average is taken over compact halo objects (in
cylindrical coordinates), and $C$ is a constant parameter. The
parameter depended on the assumptions about the form of the \pdf
and on a mean square of the eccentricity. Also,
\citet{1987ApJ...320..493L} modeled the \mw as a point. A recent
paper \citep{2010MNRAS.406..264W} offers essentially the same
method as in \citet{1981ApJ...244..805B}, and the difference lies
mainly in the arbitrary power of the radial distance in the mass
estimator. Using estimators of this kind is tantamount to
considering a very particular family of \pdf{}s. Interestingly, by
playing with various assumptions, the authors found that the \mw
mass could be as low as $4\times10^{11}\msun$ (but also as high as
$2.7\times10^{12}\msun$). \citet{1999MNRAS.310..645W} used a model
for the \mw halo with variable $\beta$. Having applied it to a
sample of satellites with known proper motions, they found the
mass within $50\kpc$ to be $5.4^{+0.2}_{-3.6}\times10^{11}\msun$
(unaffected by the presence or absence of \leoI). The limits set
on the mass within $50\kpc$ in \citet{2003A&A...397..899S} with a
larger number of tracers are similarly unaffected by \leoI:
$5.5^{+0.0}_{-0.2}\times10^{11}\msun$ (with \leoI) and
$5.4^{+0.1}_{-0.4}\times 10^{11}\msun$ (without \leoI). Mass
models considered in  \citet{2002ApJ...573..597K} give
$5.8-6.0\times10^{11}\msun$ within $100\kpc$. A more recent
estimate are $4.2\times10^{11}\msun$ within $50\kpc$
\citep{2012MNRAS.424L..44D} or $3.5-5\times10^{11}\msun$ within
$150\kpc$ for a Keplerian\footnote{Here, \textit{Keplerian} means
$M=\frac{C}{G}<r^{\gamma}v_r^2>$ with $\gamma=1$.} halo model
\citep{2012MNRAS.425.2840D} based on the same estimator as in
\citet{2010MNRAS.406..264W}. These results substantiate a low-mass
possibility within $150\kpc$.

It is interesting to ask if the mass could be reduced further and
still account for the \rvd profile. In \secref{sec:ensemble} under
spherical symmetry, we allow
 for a general \pdf, which is a
function of two integrals of motion (the energy and the
eccentricity) that describe a continuous collection of confocal
elliptical orbits. To make the approximation legitimate, we cut
the support of the \pdf off, so that the orbits' perycentra are
bounded from below by the radius of a spherical shell encompassing
the Galactic disk. Then the secondary quantities, such as the
theoretical \rvd, are obtained directly from the \pdf and can be
quite general functions of the radial variable. A particular \pdf
is found by minimizing the discrepancy between the theoretical and
the measured \rvd profiles. In this paper we do not impose
constraints on other secondary quantities, such as the density and
anisotropy profiles of the tracers. The  task of determining the
\pdf requires a good deal of numerical work. A procedure for
obtaining the \pdf is discussed briefly in \secref{sec:finding}.
In \secref{sec:ensemble} a detailed description of the theory
behind this procedure is given.

As an example of using our method in practice, we apply it to
estimate the lower bound for the \mw mass. To this end we use a
sample of tracers that are likely to be bound to the \mw if its
mass is not greater than $3.5\times10^{11}\msun$, a value that
coincides with the lowest mass estimate in the point-mass field
within $150\kpc$, as recently obtained for a Keplerian halo model
\citep{2012MNRAS.425.2840D}. This value is also consistent with
another mass estimate (without \leoI) obtained by
\citet{1999MNRAS.310..645W} for a model  with variable anisotropy
parameter and integrable halo mass. Dark halo profiles in the
literature are mostly nonintegrable. Our aim is to show that the
mass can be reduced further.

Finally, we present an example result for the same sample,
assuming the central mass of $2.4\times10^{11}\msun$. This value
is chosen for two reasons: it is greater than the lower bound we
find, and it was already obtained in the past: a) in a
three-component mass model with an asymptotic rotation velocity
$230\kms$ of the dark halo,
 fitted to the rotation of the HI layer
extending to $\approx20\kpc$  \citep{1992AJ....103.1552M}; and b)
as the best estimate at the $68\%$ confidence level obtained
assuming an isotropic velocity distribution in the point mass
field for a sample of satellites at distances of $50-140\kpc$
\citep{1987ApJ...320..493L} (where a reservation was made that the
estimate could be lower with more radial orbits).

\section{The position-velocity data}

We assume the following parameters: $R_\circ=8.5\pm0.4\kpc$ for
the Sun's distance from the Galactic center,
$V_\circ=240\pm16\km/\sec$ for the local disk rotation speed based
on three estimates ($244\pm{13}\km/\sec$ from maser data and the
motion of $Sgr A^*$ \citep{2009ApJ...704.1704B}, $V_\circ=221\pm18
\km/\sec$ from GD-1 stellar stream \citep{2010ApJ...712..260K},
and $254\pm{16}\km/\sec$ another estimate from masers). We assume
$ \br{U,V,W}=\br{11.1\pm1.7, 12.24\pm2.5, 7.25\pm0.9}\km/\sec$ for
the components of the velocity vector of the Sun with respect to
the local standard of rest based on \citet{2010MNRAS.403.1829S}
who give $U=11.1^{+0.69}_{-0.75}$, $V=12.24^{+0.47}_{-0.47}$,
$W=7.25^{+0.37}_{-0.36}$ with the additional systematic
uncertainties $(1, 2, 0.5)$. For a summary of other measurements,
see \citet{2009NewA...14..615F}.

To prepare the radial velocity dispersion profile (\rvd), which is
a sequence of averages $\disp{rv_r^2}$ over concentric spherical
shells of increasing size, we used the following position-velocity
data: the catalogs of halo giant stars
\citep{2001ApJ...555L..37D,2009ApJ...698..567S} based on the
Spaghetti Project Survey \citep{2000AJ....119.2254M}; the database
of blue horizontal branch stars \citep{2004MNRAS.352..285C} from
United Kingdom Schmidt Telescope observations and SDSS; the
database of field horizontal branch and A-type stars
\citep{1999AJ....117.2329W} based on the survey of
\citet{1992AJ....103.1987B}; the catalogs of globular clusters
\citep{1996AJ....112.1487H} and dwarf galaxies
\citep{1998ARA&A..36..435M}. The data was recalculated to epoch
J2000 when necessary. In addition, we included the ultra-faint
dwarf galaxies such as \textit{Ursa Major I} and \textit{II},
\textit{Coma Berenices}, \textit{Canes Venatici I} and
\textit{II}, \textit{Hercules} \citep{2007ApJ...670..313S},
\textit{Bootes I}, \textit{Willman 1} \citep{2007MNRAS.380..281M},
\textit{Bootes II} \citep{2009ApJ...690..453K}, \textit{Leo V}
\citep{2008ApJ...686L..83B}, \textit{Segue I}
\citep{2009ApJ...692.1464G}, and \textit{Segue II}
\citep{2009MNRAS.397.1748B}. To eliminate a possible decrease in
the \rvd at smaller radii due to circular orbits in the disk, we
excluded tracers in the ellipsoidal disk vicinity:
$\frac{Z^2}{a^2}+\frac{R^2}{b^2}\,\leq1$ with $a=4\kpc$ and
$b=20\kpc$. Here, $20\kpc$ agrees with the extension of the
observable disk \citep{1984PASP...96..841C}, while $4\kpc$ at
smaller radii agrees with a criterion used by
\citet{2008ApJ...684.1143X} to exclude some thick-disk stars for
which $|Z|<4\kpc$. We excluded \textit{Leo T} located at
$r>400\kpc$ (its large spatial separation from closer tracers
makes it unsuitable for preparing the \rvd profile).

\medskip

There is a large uncertainty in the \mw mass determination,
especially at large distances $\ga 50\kpc$, mainly due to the
poorly constrained spatial extent of the dark halo, dominating
baryonic mass components. Total mass estimates can differ by a
factor of $4$ or more, ranging from $0.5\times 10^{12}\msun$ to
$2\times 10^{12}\msun$ \citep{2010AJ....139...59B} and more. The
dark halo's status is entirely speculative
\citep{1998MNRAS.294..429D}. As a result, it is not a priori known
which of high-velocity tracers are gravitationally bound to the
\mw. Mass estimates of the \mw are largely affected by several
such tracers \citep{2003A&A...397..899S}. Apart from \leoI these
are: Pal 3, Draco, and a few FHB stars. However, \leoI seems
almost certainly unbound, as the argument below shows.

\subsection{The case of \leoI}

\begin{figure}
\centering
\includegraphics[width=0.5\textwidth]{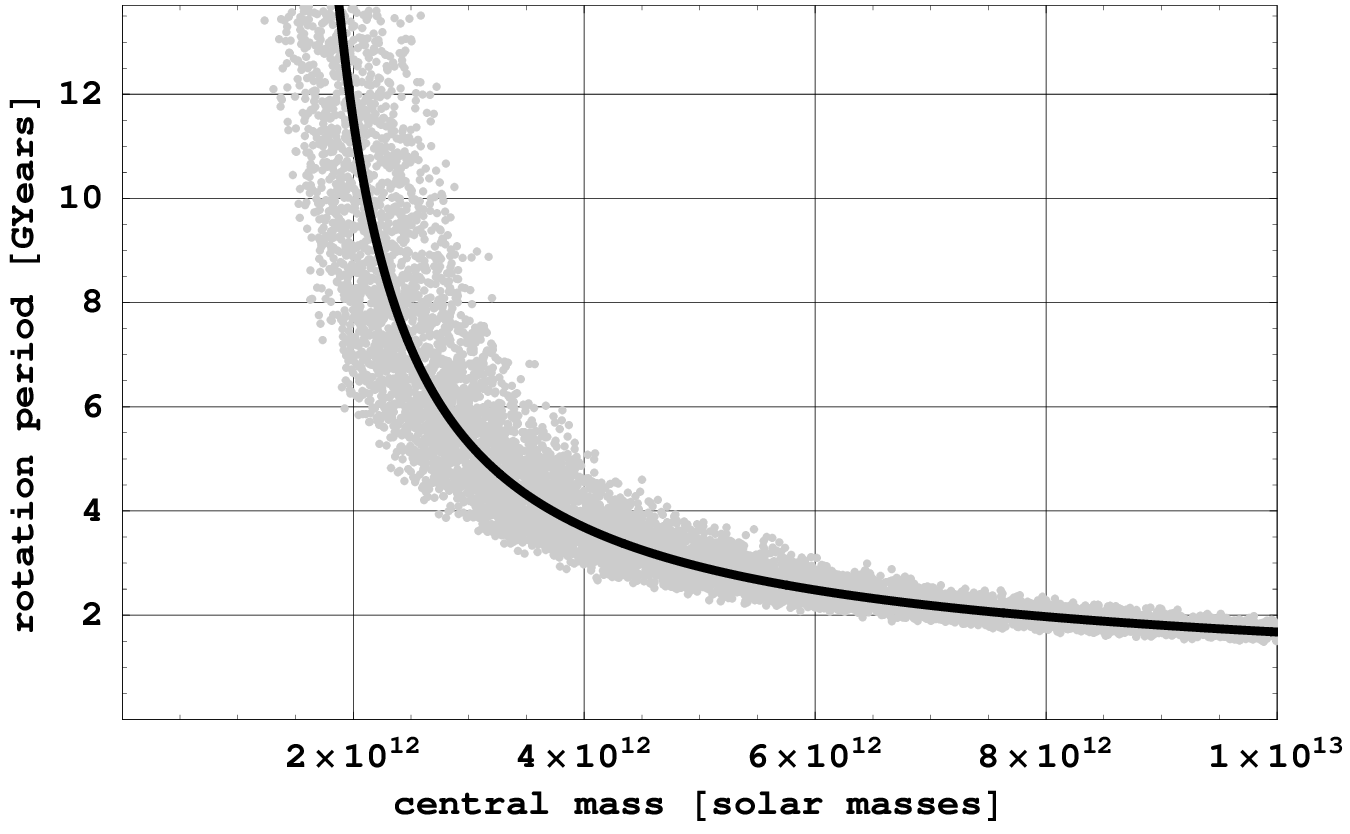}\\
\includegraphics[width=0.5\textwidth]{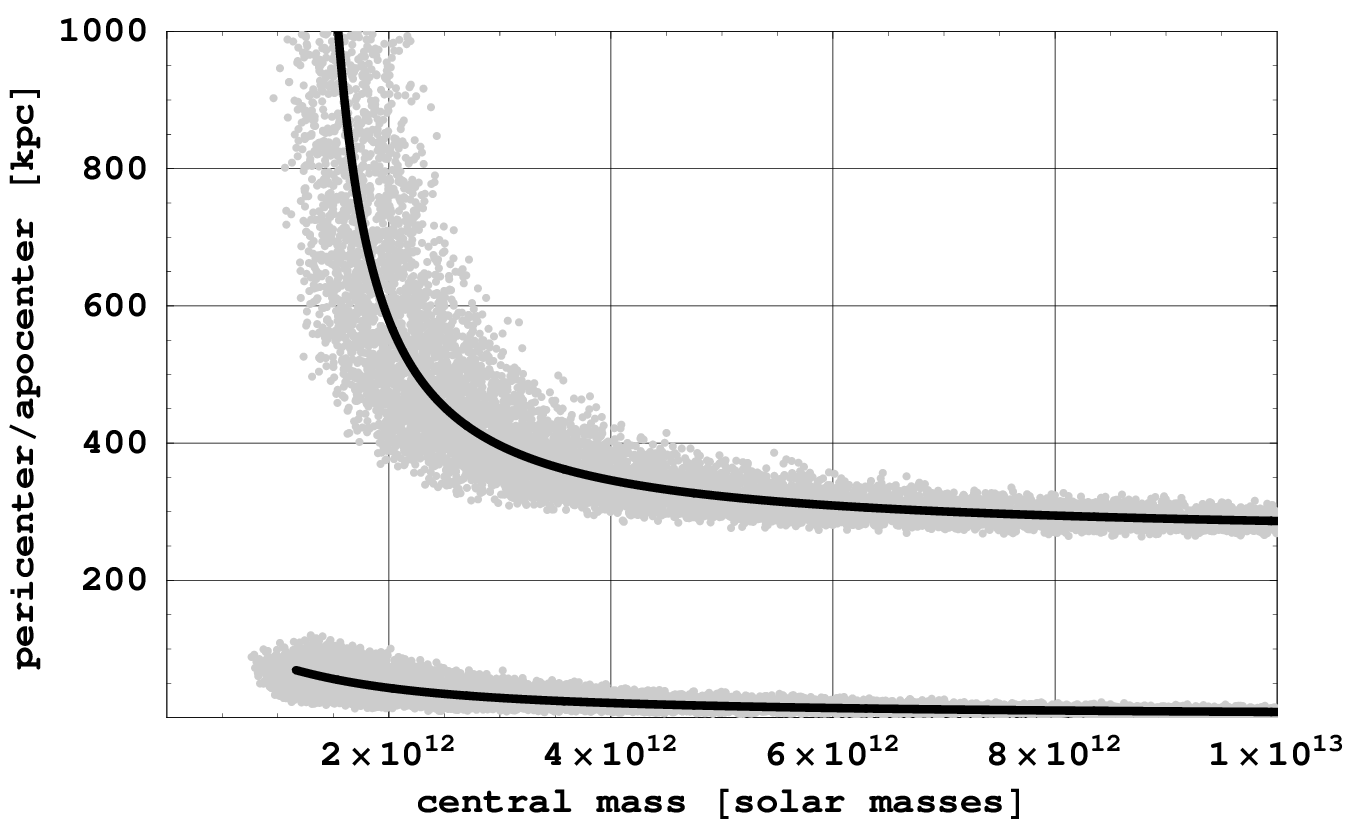}\\
\caption{\label{fig:LeoI} The orbital period, pericenter, and
apocenter for \leoI shown as functions of the central mass in the
point mass potential, calculated according to \leoI's
Galactocentric position and velocity given in
\citep{2013ApJ...768..139S}. The gray dots represent random values
for bound (elliptic) orbits admissible within errors.
}\end{figure}

\leoI is a dwarf spheroidal galaxy, a very distant and
fast-receding satellite of the \mw. As follows from the escape
velocity argument, \leoI could be bound to the \mw, if \mw's mass
was greater than
$\frac{1}{2G}r(v_{\mathrm{rad}}^2+v_{\mathrm{tan}}^2)=
1.16\times10^{12}\msun$, where we have used values given in
\citet{2013ApJ...768..139S}. As can be seen in \figref{fig:LeoI},
the central mass should be much higher than $2\times10^{12}\msun$,
if one agrees that the orbital period of a relaxed bound object
should be much lower than the Hubble time. (\leoI is often
included in the Jeans analysis which by definition assumes a
stationary relaxed system of tracers.) Furthermore, it would be
unreasonable to expect the apocenter of a \mw's satellite (that
is, a body not affected gravitationally by the presence of other
neighboring high-mass concentrations, like M31) to be close to, or
further than halfway distance between large galaxies in the Local
Group, which is $\approx350\kpc$. This gives at least
$3\times10^{12}\msun$ for the \mw mass, as seen from
\figref{fig:LeoI}. Also, the timing argument
\citep{1959ApJ...130..705K} leads to a comparable mass
$\mvir=3.15^{+1.58}_{-1.36}\times10^{12}\msun$
\citep{2013ApJ...768..139S}. But a timing argument is likely to
overestimate the true mass, as \citet{2008ApJ...678..187V} noticed
for the M31-\mw system, even by a factor of $2$. Such high mass is
also improbable according to the argument by
\citet{2012MNRAS.424.2715W}: the probability that \mw's \lcdm halo
has the observed number of subhalos with a given maximum circular
velocity, decreases steeply with increasing \mw's virial mass,
effectively vanishes for $\mvir>3\times10^{12}\msun$, and is only
$5\%$ for $\mvir>2\times10^{12}\msun$. Therefore, the \mw mass
cannot be too high, so the timescale for \leoI must be within an
order of magnitude of the Hubble time. This contradicts the
assumption that the time scale is much lower.

High masses from the timing argument are also inconsistent with
recent mass estimates. The favored model in
\citep{2002ApJ...573..597K} gives $10^{12}\msun$ for the virial
mass of the \mw. \citet{2008ApJ...684.1143X} estimate the \mw mass
to $4.0\pm0.7\times10^{11}\msun$ within $60\kpc$ from the
kinematics of a large virialized sample of BHB stars, and
extrapolate this result to the \mw's dark halo mass
$\mvir=10^{+3}_{-2}\times10^{11}$. \citet{2012MNRAS.425.2840D}
suggest that the mass within $150\kpc$ is probably in the range
from $5.0\times10^{11}\msun$ to $10.0\times10^{11}\msun$ and that
there may be low mass between $50$ and $150$ $\kpc$ (implying a
high-concentration halo). The authors come to the conclusion, that
\leoI is almost certainly unbound.  Next,
\citet{2013MNRAS.431.1220D} conclude with the \mw's mass is
between $5.5-7.5\times10^{11}\msun$. Another recent work
\citep{2013MNRAS.428.1696V} finds that the number and internal
dynamics of the classical dwarf spheroidal satellites will be
consistent with the predictions of the \lcdm model, if the \mw
total mass is $8.0\times10^{11}\msun$, again insufficient to bind
\leoI to the \mw. The latter value was obtained in the past as the
upper bound for the mass
 at the $99\%$ confidence level, by
assuming an isotropic velocity distribution for objects at
distances of $50-140\kpc$ \citep{1987ApJ...320..493L}, with the
reservation that this estimate can be lower for more radial
orbits.

\leoI is a kinematical outlier from the rest of velocity tracers.
In this context we recall that the statistical analysis of data
advises not including single measurements that evidently are well
outside the range of other data. It would hardly be an acceptable
situation if the \mw mass was determined using a single tracer
(while it is uncertain whether it is bound or not). By the mere
assumption that \leoI is bound, one would have to accept a lower
limit for the \mw mass of at least $1.16\times10^{12}$, even
though motions of a vast majority of remaining tracers or other
arguments as in \citet{2012MNRAS.424.2715W} might point to a lower
mass, insufficient to bind \leoI. As tested by
\citet{2011MNRAS.415.2607D}, including unbound satellites in two
popular mass estimates
\citep{1981ApJ...244..805B,2010MNRAS.406..264W} can cause large
overestimates of the true mass. \leoI disproportionately affects
\mw mass estimates under the assumption of equilibrium kinematics
\citep{2013ApJ...768..139S}, e.g. by adding \leoI, the mass
estimate can be increased by nearly a factor of $3$
\citep{1989ApJ...345..759Z}. A simple mass estimator applied to
the satellite populations (\leoI discarded)  gives the minimum
$4\pm1\times10^{11}\msun$ for enclosed mass within $300\kpc$
\citep{2010MNRAS.406..264W}, whereas including \leoI would
increase the estimate to $15.0\pm4.0\times10^{11}\msun$. By using
a halo model \citep{1999MNRAS.310..645W} with variable anisotropy
parameter and integrable halo mass, one can show that both the
mass and the length scale change by a lot: when \leoI is included
the most likely values are $17\times10^{11}\msun$ and $150\kpc$
for the total halo mass
 and the length scale, whereas with \leoI
excluded, the quantities shrink by a factor of $6$ to
$3.0\times10^{11}\msun$ and $25 \kpc$. According to us,
determination of mass should be stable against inclusion/exclusion
of a sufficiently small subsamples of velocity tracers. With \leoI
this is by no means possible.

As it was suggested in \citet{2013ApJ...768..139S}, there might be
a $77\%$ chance that \leoI could be bound to the \mw. However, the
same authors conclude that it would not necessarily be appropriate
to include \leoI in equilibrium models used to estimate the \mw
virial mass. The kinematics of \leoI is unlikely to be virialized
on account of its expected first infall into the \mw. It was
concluded in \citet{1994AJ....107.2055B}, that \textit{the history
of the local group is too complex to justify calculating the mass
of the \mw by assuming that all satellites of our galaxy are
bound}. Second, a more prosaic scenario cannot be excluded: \leoI
may have passed coincidentally by us, and could have originated
outside the \mw \citep{1989ApJ...345..759Z}. Third, as observed in
\citet{2007MNRAS.379.1475S}, fast receding satellites, such as
\leoI, can be present owing to a three-body ejection mechanism,
which is propelling bodies into highly energetic orbits.

Given the above arguments, it seems most likely  that \leoI must
be a member of a higher mass concentration and cannot be bound to
the \mw alone. We therefore discard \leoI from further analysis.

\subsection{\label{sec:TrivialEstimator}Two samples of tracers}

\begin{figure}
\hspace{-0.04\textwidth}
\includegraphics[width=0.56\textwidth]{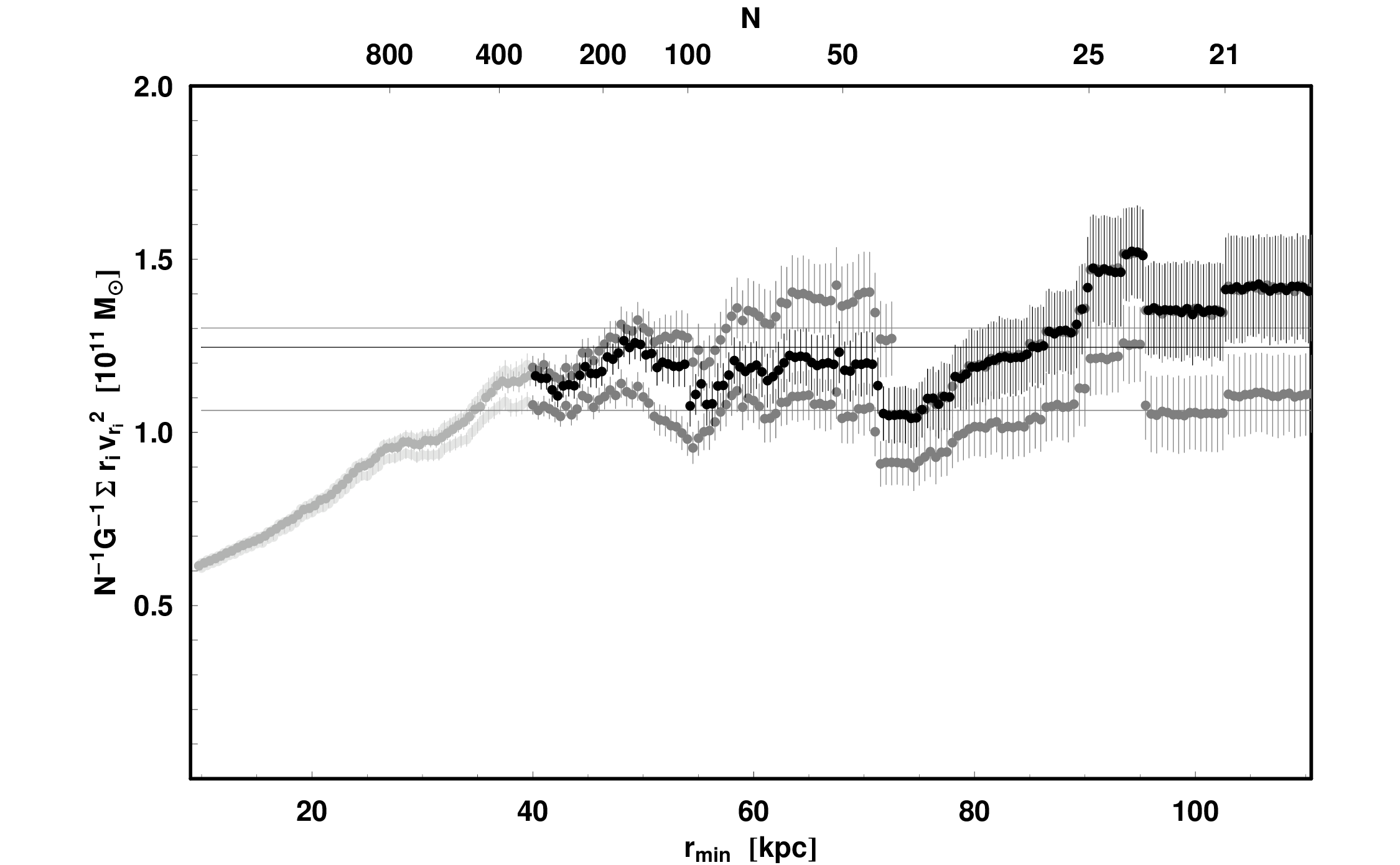}\\
\caption{\label{fig:MassEstimate} The mean values $\frac{1}{N
G}\sum_i r_i\,v_{r,i}^2$; $r_i>r_{\mathrm{min}}$, $N=\sum_i 1$;
for tracers outside the spherical surface of radius $r_{min}$
shown as a function of that radius (the vertical bars represent
the standard deviation for random subsets with $7/8$ of the
elements). The curve in the top \textit{[top, gray]} (the \tII
sample) discards only \leoI; the curve in the middle
\textit{[middle, black]} discards an additional single tracer
$J160826.42+065542.3$ with
$\frac{1}{2G}rv_r^2>5.0\times10^{11}\msun$; the curve in the
bottom \textit{[bottom, gray]} (the \tI sample), in addition to
the previous two tracers, discards more $4$ tracers
(\textit{88-TARG37}, \textit{Hercules},
\textit{J234809.03-010737.6}, and \textit{J124721.34+384157.9})
with $\frac{1}{2G}rv_r^2\geq 3.49\times10^{11}\msun$. The two
lower curves are approximately flat in a wide range of radii
$40\kpc<r_{min}<90 \kpc$. The thin straight lines show the mean
values in that region. It is important to note that the resulting
mass estimate for sample \tII compared to that for sample \tI is
increased by only a factor of $\approx1.16$, which is comparable
to or even less than model errors for the \mw mass in the
literature.}\end{figure}

When the \mw's mass is expected to be lower, as suggested by the
discussion in \secref{sec:intro}, some of the tracers cannot be
gravitationally bound to it and, therefore, similar to \leoI,
should not be included in preparing the \rvd profile. The
following criterion can be used for hypothesizing which of
satellites might be unbound. The simple calculation of
\secref{sec:intro} leads to the mass estimator
$\widetilde{M}_r=\frac{\mu }{N_r}\sum_{r_i<r} r_i\,v_{r,i}^2$ with
$\mu=4$ for $\beta\approx0$ (or $\beta\approx\epsilon/2$ with a
small $\epsilon>0$), where the summation is taken over a number
$N_r$ of objects with the radial distance not lower than $r$. This
result follows at once for an approximately flat $r\disp{v_r^2}$,
when one can use $\disp{rv_r^2}$ instead, but it can also be
arrived at without this reservation by applying Eq.23 given in
\citet{2010MNRAS.406..264W} with suitable parameters. As can be
seen in \figref{fig:MassEstimate},
the mass estimate $\widetilde{M}\approx 4-5\times10^{11}\msun$
($\mu\approx4$) is approximately invariable in the region from
$40\kpc$ to $90\kpc$, consistently with what one would expect for
a compact concentration of mass. (Beyond $90\kpc$ the statistics
is too small.) This is in accord with the value
$3.5-5\times10^{11}\msun$ within $150\kpc$, obtained in a
Keplerian halo model \citep{2012MNRAS.425.2840D}. With the
limiting value $\mu=2$ for $\beta=1$, $\widetilde{M}$ might be as
low as $2.1-2.6\times10^{11}\msun$ (see, the straight lines in
\figref{fig:MassEstimate}). The two estimates suggest that
satellites with $\frac{1}{2G}r v_r^2$ greater than
$5\times10^{11}\msun$ or $3\times10^{11}\msun$, respectively,
might not be bound to the \mw and should be excluded prior to
preparing the \rvd profile. We further assume these two
possibilities by considering two samples of tracers \tII and \tI
as specified in Table \ref{tab:I}.
\begin{table}
\centering
\begin{tabular}{c|c|c|@{}c|c|@{}c}
\hline\hline
tracer's & short & ${rv_r^2}/{2G}$ & distance &  &  \\
name & name & $[10^{11}\msun]$ & $[\kpc]$ & \tII & \tI  \\
\hline\hline \leoI &  & $8.29$ & $254$ & $-$ & $-$ \\
\hline
\textit{J160826.42+065542.3} & \tX & $5.62$ & $72.7$ & $-$ & $-$\\
\hline
\textit{88-TARG37} & \tA & $4.00$ & $55.3$ & $+$ & $-$ \\
\textit{Hercules} & \tB & $3.65$ & $132$ & $+$ & $-$ \\
\textit{J234809.03-010737.6} & \tC & $3.51$ & $54.2$ & $+$ & $-$  \\
\textit{J124721.34+384157.9} & \tD & $3.49$ & $41.5$ & $+$ & $-$  \\
 \hline
\textit{J232526.89-094433.5} &  & $2.88$ & $53.8$ & $+$ & $+$  \\
\textit{J121222.29+422502.0} &  & $2.84$ & $37.7$ & $+$ & $+$  \\
\textit{J115555.42+365908.6} &  & $2.77$ & $51.0$ & $+$ & $+$  \\
\textit{J090016.24+341342.4} &  & $2.61$ & $48.8$ & $+$ & $+$  \\
\textit{J144258.28+575339.3} &  & $2.57$ & $48.3$ & $+$ & $+$  \\
\textit{Eridanus} &  & $2.56$ & $95.4$ & $+$ & $+$  \\
\textit{J160507.08+293222.3} &  & $2.54$ & $42.2$ & $+$ & $+$  \\
\textit{Pal14} &  & $2.52$ & $71.3$ & $+$ & $+$  \\
\textit{85-TARG17} &  & $2.34$ & $45.1$ & $+$ & $+$  \\
$\vdots$ &  & $\vdots$ & $\vdots$ & $\vdots$ & $\vdots$ \\
\hline\hline
\end{tabular}
\caption{\label{tab:I} The radial velocity tracers with largest
$\frac{rv_r^2}{2G}$ and samples \tII and \tI used in the text. The
numerical values were calculated in the Galactocentric coordinate
system, assuming $V_o=240\kms$, $R_o=8.5\kpc$, $U=11.1\kms$,
$V=12.24\kms$, and $W=7.25\kms$.}
\end{table}

\subsection{\label{sec:rvdprofile}The radial velocity dispersion profile}

\begin{figure}
\begin{tabular}{@{}c@{}}
\hspace{-0.015\textwidth}
\includegraphics[width=0.50\textwidth]{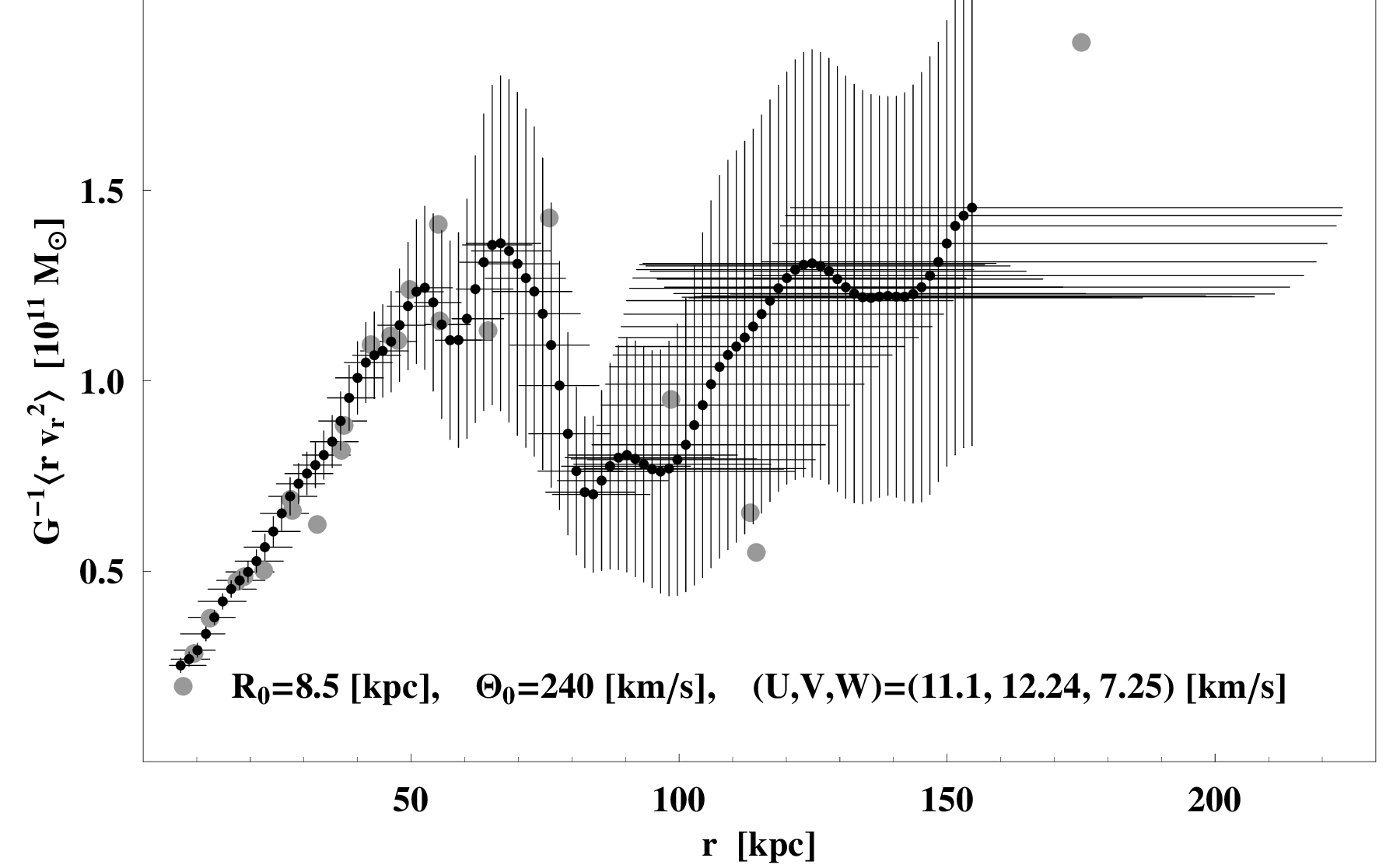}\\
sample \tII\\
\hline {}\\\hspace{-0.015\textwidth}
\includegraphics[width=0.50\textwidth]{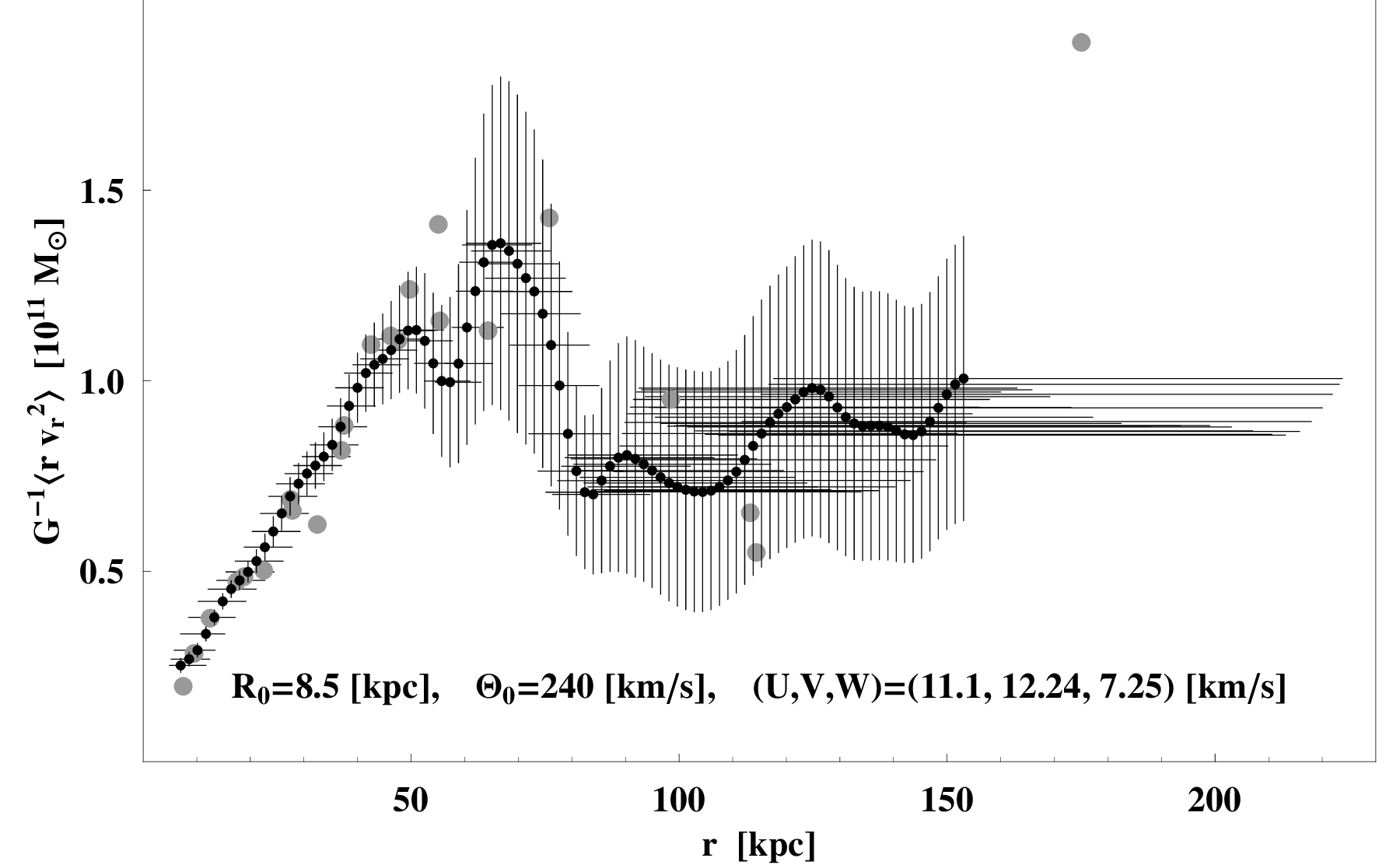}\\
sample \tI\\
\hline
\end{tabular} \caption{\label{fig:RVDprofiles} Radial velocity
dispersion (\rvd) profiles $G^{-1}\disp{rv_r^2}$  for tracers with
$\frac{1}{2G}rv_r^2<5.0\times10^{11}\msun$ (the \tII sample)
\textit{[top panel]}; and
$\frac{1}{2G}rv_r^2<3.49\times10^{11}\msun$ (the \tI sample)
\textit{[bottom panel]}. The horizontal bars represent the
effective radial bin size of the moving average. The vertical bars
indicate the spread in the \rvd due to the inclusion/exclusion of
random subsets of tracers. A detailed description of how these
profiles were obtained can be found in \secref{sec:rvdprofile}. As
reference values, we show $G^{-1}r\disp{v_r^2}$ calculated based
on the \rvd points in \citep{2008ApJ...684.1143X} and
\citep{2012MNRAS.425.2840D} \textit{[large gray circles]} (note
that \leoI was excluded from our analysis). See also the
comparison of the profiles in \figref{fig:RVDBackgr}}
\end{figure}

\begin{figure}
\hspace{-0.005\textwidth}
\includegraphics[width=0.50\textwidth]{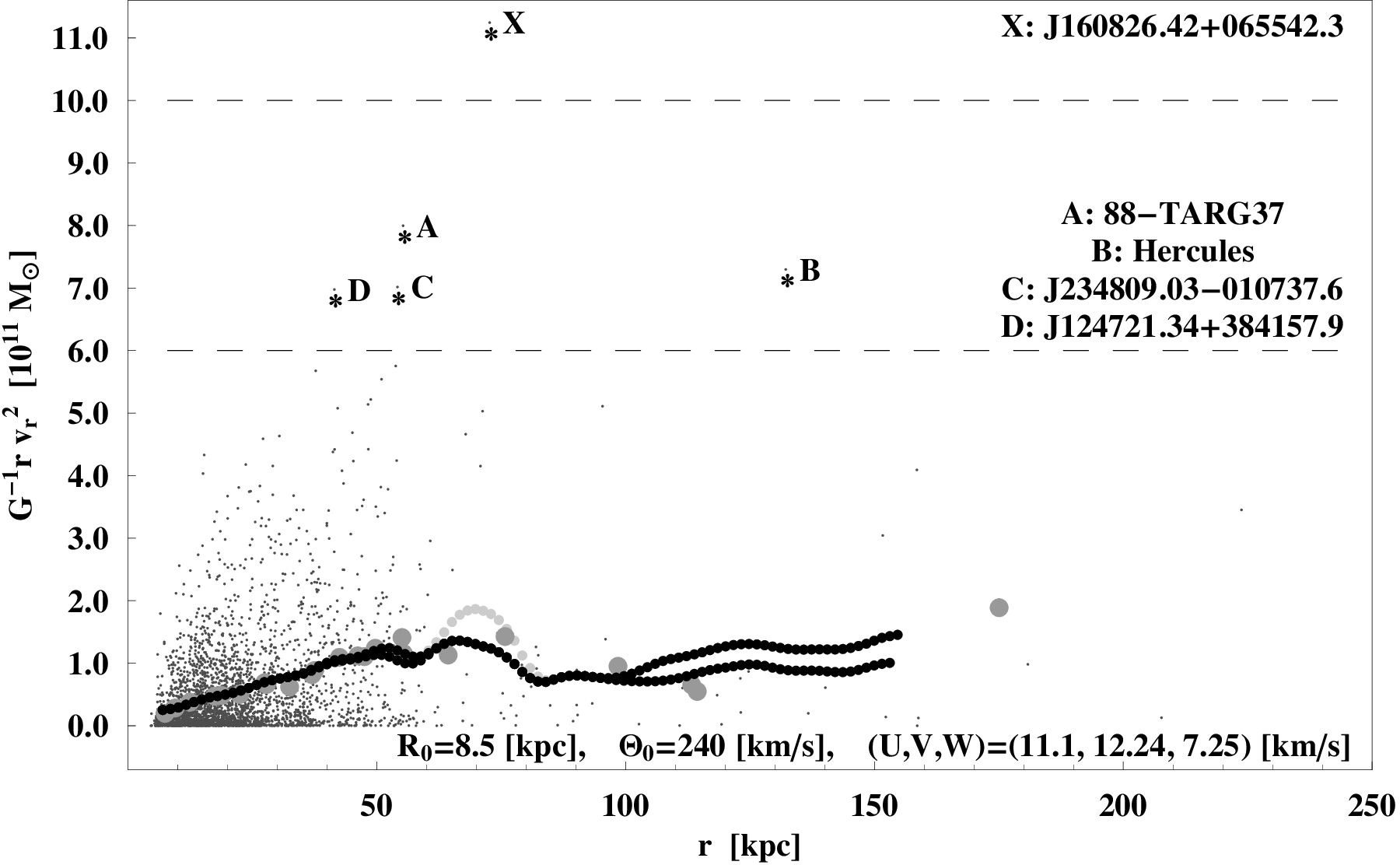}\\
\caption{\label{fig:RVDBackgr} The radial velocity tracers
\textit{[gray dots/black stars]} and the radial velocity
dispersion (\rvd) profiles \textit{[black circles]} of
\figref{fig:RVDprofiles} shown on the $\br{r,G^{-1}rv_r^2}$ plane.
The \rvd{}s are curves prepared assuming that either only \leoI
and \tX (the \tII sample) or both these two and \tA, \tB, \tC, \tD
(the \tI sample), are excluded (see Table \ref{tab:I}). With all
tracers included (with the exception of \leoI), a bump structure
centered at $\approx70\kpc$ would appear \textit{[light gray
circles]}. The inclusion or exclusion of \tA, \tC, and \tD does
not influence the \rvd{}s significantly (only a little bit close
to $\approx50\kpc$), whereas the inclusion of \tB rises the \rvd
curve for radii greater than $100\kpc$. Thus, the \rvd curves are
mainly dependent only on two tracers \tB (\textit{Hercules}) and
\tX. The \textit{large gray circles} are the reference
$G^{-1}r\disp{v_r^2}$ values based on the literature (the same as
in \figref{fig:RVDprofiles}). }
\end{figure}

As explained in \secref{sec:TrivialEstimator}, we consider two
\rvd profiles based on tracers from samples \tII and \tI (see
Table \ref{tab:I}). The \pdf{}s are shown in
\figref{fig:RVDprofiles}
and also compared in \figref{fig:RVDBackgr}, in the background of
all tracers.

The \rvd curves in \figref{fig:RVDprofiles} were obtained as
follows: for any $r_i$
 all of $n_i$ tracers are taken inside a spherical shell
$|r-r_i|<w/2$ of a fixed width $w$. When necessary, $w$ is
increased so that $n_i$ is never lower than some fixed $n$, which
effectively increases the bin size at large radii where the
statistics are poor because there were few tracers. Then $m$
random subsets of size $\frac{2}{3}n_i$ are chosen, and a quartet
of numbers is assigned to each of them: two mean values $\disp{r}$
and $G^{-1}\disp{rv_r^2}$, and two numbers $r_{min}$ and
$r_{max}$, the smallest and the largest $r$ in each subset. This
is repeated for all $r_i$ at various $n$'s. Then an ordered list
consisting of all the quartets is formed, sorted according to
increasing $\disp{r}$. Finally, a moving average is performed over
quartets with $\disp{r}$ falling in a window of width $W$, when
the window moves through the entire range of $r$. In effect, a
curve is obtained on the plane $({r},G^{-1}{rv_r^2})$. We assumed
$w=9\kpc$, $m=89$, $W=6\kpc$, and $n$ from $15$ to $22$. For each
$r$ the horizontal ``error bars'' represent the intervals
$\disp{r_{min}}<r<\disp{r_{max}}$, while the vertical ``error
bars'' represent the standard deviations of $G^{-1}\disp{rv_r^2}$
times $\sqrt{3}$, and measure a Monte Carlo-like estimate of the
uncertainty in $G^{-1}\disp{rv_r^2}$ due to various subsets of
tracers taken in obtaining the mean values $\disp{rv_r^2}$.
%

It can be seen that the \rvd profiles largely depend on two
tracers with the highest $\frac{1}{2G}rv_r^2$, \tB, and \tX, while
the inclusion/exclusion of the other three tracers \tA, \tC, and
\tD, is not so decisive. Tracer \tX certainly cannot be bound to
the \mw due to the recent low-mass estimates referred to earlier
and therefore should not be included. This consideration
substantiates our choice of the two basic \rvd profiles.

The \rvd profiles in Fig. \ref{fig:RVDprofiles} can be
interpolated to form continuous curves that can be regarded as
smooth model curves, consistent with the measurements within
certain uncertainty limits.\footnote{If need be, one can choose a
subset of points on these curves with the respective ``error
bars'' akin to the \rvd data consisting of few points as usually
given in the literature, such that the difference between the
abscissas of the neighboring points are determined by the
extension of the respective horizontal bars (e.g., a half of the
mean of the extension). There are many such subsets, depending on
the position of the initial point; however, all these subsets will
be consistent with measurements within ``errors''.} With the aid
of the method introduced in \secref{sec:finding}, a number of
\pdf{}s can be found, giving rise to the respective theoretical
\rvd curves consistent with the smooth model curves within the
same limits. When this is possible with some mass $M$, we say that
the $M$, as an estimate for the \mw mass, accounts for the radial
motions of tracers.

\section{\label{sec:ensemble} The Keplerian ensemble method}

To estimate the lower bound for the \mw mass, we assume that
external tracers move as test bodies in the host gravitational
field of a compact mass distribution. In this case, at
sufficiently large radii, a contribution from the higher
multipoles of the field should be small compared to the monopole
part, and the motion of distant test bodies should be
approximately describable in terms of Keplerian orbits. Because
point mass formulae can be used as order-of-magnitude estimates of
gravitating mass, the compactness assumption does not necessarily
reject the possibility that an extended massive dark matter halo
may be present. Also, the compactness assumption does not a priori
reject the possibility that the spatially extended dark matter
halo component may be absent. The latter hypothesis should be
tested in accordance with Occam's razor rule, before one decides
to introduce new forms of matter, such as a ubiquitous invisible
substance consisting of dark matter particles of unknown nature
and, so far, eluding detection in the laboratory by producing no
observable non-gravitational effects (e.g., recent null results,
\citet{2013arXiv1310.8214L}).

\subsection{The method} The motion of a test body in a spherically
symmetric potential $\Phi(r)$ is flat. It occurs in a plane
through the center of symmetry. The plane is determined by two
integrals of motion that fix the unit vector normal to the plane.
The additional two integrals are energy $E$ and the magnitude of
angular momentum $J$ (both per unit mass). In terms of angular
velocity $\omega$,
$\omega^2=\dot{\theta}^2+\dot{\phi}^2\sin^2{\theta}$, they read as
$r^2\omega^2=\frac{J^2}{r^2}$ and $
v_r^2=2E-2\Phi-\frac{J^2}{r^2}$. In the special case of Newtonian
potential $\Phi=-\frac{GM}{r}$, there is an additional integral of
motion indicating a fixed direction in the plane of motion.

The condition $v_r^2\geq0$ for all $r>0$ with $J\neq0$ gives
$1+\frac{2E J^2}{G^2M^2}\geq0$. We are only interested in
spatially bound orbits. Two turning points (where $v_r=0$) will be
possible for $E<0$ and $1+\frac{2E J^2}{G^2M^2}<1$. Elliptical
orbits are therefore possible only for pairs $(E,J)$ such that
$0\leq 1+\frac{2E J^2}{G^2M^2}<1$. Elliptical motion can be thus
uniquely determined by specifying five numbers: three Euler angles
describing orientation of the orbit in space, and a pair of
numbers $(e,\e)$ defined by $e^2={1+\frac{2E J^2}{G^2M^2}}$,
$0\leq e<1$, and $\e=-\frac{R E}{GM}>0$. Here, $R$ is an arbitrary
unit of length, $e$ the eccentricity, and $\e$ a measure of energy
determining the length of the large semiaxis, which is
$\frac{R}{2\e}$. The turning points are $\frac{R}{2\e}\br{1\pm
e}$. The dimensionless parameters $(e,\e)$ play the central role
in further considerations.

To find various expectation values for a spherically symmetric
collection of confocal ellipses (called a Keplerian ensemble), it
suffices to know a \pdf that describes the number of ellipses with
various $e$ and $\e$. It will be related to the distribution
function $f(\vec{r},\vec{v})$ in $\mu$-phase space. We assume that
$f$ can be expressed through first integrals $e$ and $\e$. Then,
$f$ is stationary and satisfies the necessary condition
$\partial_tf+\partial_{q^i}f+\partial_{\dot{q}^i}f\partial^{i}\Phi=0$
for a collisionless system. This is the wording of the Jeans
theorem (\citealt{1915MNRAS..76...70J}) in our situation. In
general, the theorem states that for a system in steady state, the
\pdf is a function of isolating integrals of motion
(\citealt{1915MNRAS..76...70J}). (Under spherical symmetry two
such integrals suffice).

\subsubsection{A cutoff phase space and the resulting expectation
values}\label{sec:expect_val}

The integrals of motion $e$ and $\e$ can be regarded as new,
independent phase variables. By making a transformation from
ordinary spherical coordinates
$r,\theta,\phi,v_r,v_{\theta},v_{\phi}$ to new coordinates
$u,\theta,\phi,\e,e,\psi$ of the form $r\to{}R\,u$,
$\br{\theta,\phi}\to\br{\theta,\phi}$,
$v_r^2\to\frac{GM}{R}\br{\frac{2}{u}-
\frac{1-e^2}{2\e}\frac{1}{u^2}-2\e}$, and
$\br{v_{\theta},v_{\phi}}
\to\sqrt{\frac{GM}{R}}u^{-1}{\br{\frac{1-e^2}{2\e}}^{1/2}}\br{\sin{\psi},\cos{\psi}}$,
the original volume element
$r^2\ud{r}\,\sin{\theta}\,\ud{\theta}\,\ud{\phi}\,
\frac{\ud{v_r^2}}{2v_r}\,\ud{v_{\theta}}\,\ud{v_{\phi}}$ is
transformed (up to a constant factor) to
$\mathcal{J}(e,\e,u)\,\ud{u}\,\sin{\theta}\,\ud{\theta}\,\ud{\phi}\,
\ud{\psi}\,\ud{\e}\, e\,\ud{e}$, with
$$\mathcal{J}(e,\e,u)=\sq{\e\br{\e-\frac{1-e}{2u}}\br{\frac{1+e}{2u}-\e}}^{-1/2}.$$
Since we assume spherical symmetry, the angles $\phi$, $\theta$,
and $\psi$ can be integrated out. The remaining part of the
integration domain is determined by the function $\mathcal{J}$.
Furthermore, in finding the expectation values as functions of
$r$, the integration must be taken over all ellipses that
intersect a spherical thin shell of a fixed radius $r$, concentric
to the center of symmetry. Given $0<r<\infty$ and $0\leq e<1$, we
have $\frac{1-e}{2u}\leq \e \leq\frac{1+e}{2u}$ for such orbits.
Hence, we arrive at the distribution integral
\begin{equation}\label{eq:origphaseintegral}\int_{0}^{\infty}\!\ud{u}\,
\int_{0}^{1}e\,\ud{e}
\int_{\frac{1-e}{2u}}^{\frac{1+e}{2u}}\!
\ud{\e}\,\mathcal{J}(e,\e,u)f\br{e,\e},\end{equation} which equals
$\int f(\vec{r},\vec{v})\ud{}^3{\vec{r}}\ud{}^3{\vec{v}}$ to
within a constant factor.

\begin{figure}
\centering
\includegraphics[width=0.4875\textwidth]{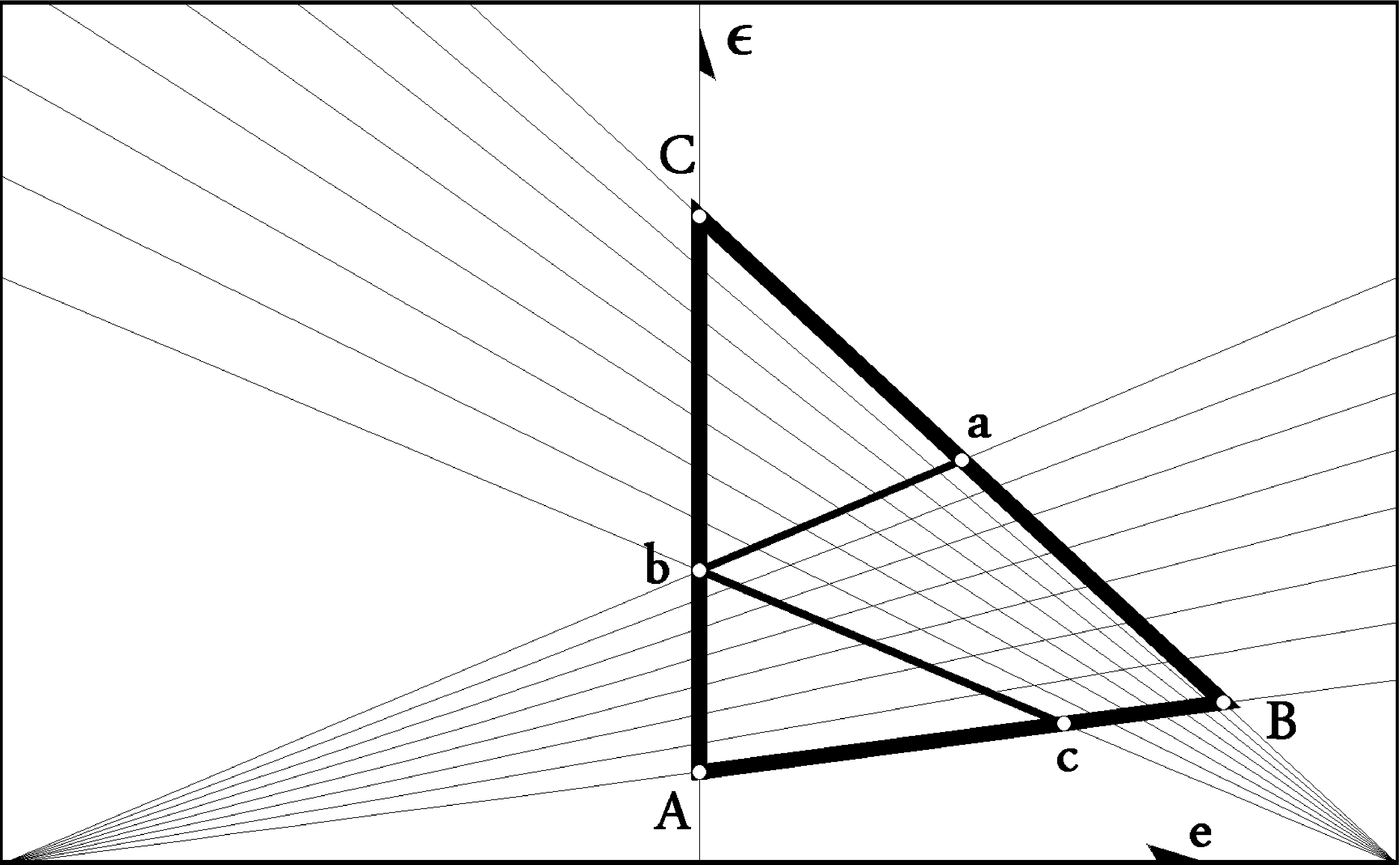}\\
\caption{\label{fig:support} Each point on the $(e,\e)$ plane
represents an elliptical orbit with eccentricity $e$ and energy
$-\e$. The \textbf{ABC} triangle with vertices \textbf{A}
$(0,\frac{1}{2\B})$, \textbf{B}
$\br{\frac{\B-\A}{\B+\A},\frac{1}{\A+\B}}$, \textbf{C}
$\br{0,\frac{1}{2\A}}$ is a locus of all confocal ellipses lying
entirely in between two bounding spheres of fixed radii $\A$ and
$\B$ and centered on the focal point. The two families of mutually
crossing lines $\e=\frac{1+e}{2u}\alpha$ through point $(-1,0)$
and $\e=\frac{1-e}{2u}\beta$  through point $(1,0)$ give rise to a
new coordinate system $(\alpha,\beta)$ on the $(e,\epsilon)$
plane.  When $\alpha\in\br{\frac{u}{\B},1}$ and
$\beta\in\br{1,\frac{u}{\A}}$ the coordinates cover a
$u$-dependent quadrilateral integration domain \textbf{abcB} with
vertices \textbf{a} $\br{\frac{u-\A}{u+\A},\frac{1}{u+\A}}$,
\textbf{b} $\br{0,\frac{1}{2u}}$, \textbf{c}
$\br{\frac{\B-u}{\B+u},\frac{1}{\B+u}}$, and \textbf{B}, which is
the locus of all orbits $(e,\e)$ crossing at least once a sphere
of a given radius $\A<u<\B$.}
\end{figure}

Next, for the physical reasons, we assume all orbits of the
ensemble to be contained entirely within a spherical shell
$R\,\A<r<R\,\B$, that is, in between two boundary spheres of radii
$r_a=R\A$ and $r_b=R\B$. (As a byproduct, the normalized
cumulative number of objects will be automatically integrable.)
This spatial boundary imposes additional restrictions on the
integration domain in the phase space, changing considerably the
support of $f(\e,e)$. Then, $\A<\frac{1-e}{2\e}$ and
$\frac{1+e}{2\e}<\B$. Hence, $\frac{1+e}{2\B}<\e<\frac{1-e}{2\A}$
and also $0\leq e<\frac{\B-\A}{\B+\A}<1$. As a result, the
integration domain in the integral \eqref{eq:origphaseintegral}
gets shrunk to a quadrilateral \textbf{abcB}, as shown in
\figref{fig:support}.
However, the integration requires splitting the $u$-dependent
domain into three parts. To overcome this difficulty we introduce
a mapping $(\alpha,\beta)\to(e(\alpha,\beta),\e_u(\alpha,\beta))$
from a rectilinear region with cartesian coordinates
$(\alpha,\beta)$ to a $u$-dependent region \textbf{abcB}. This
leads to the following coordinate change in the reduced phase
space $(\e,e,u)$ in the above distribution integral:
\begin{eqnarray*}&e\to
e\br{\alpha,\beta}=\frac{\beta-\alpha}{\beta+\alpha}, \quad
\e\to\e
\br{u,\alpha,\beta}=\frac{1}{u}\frac{\alpha\beta}{\alpha+\beta},
\quad u\to u,&\\ &\frac{u}{\B}<\alpha<1,\quad
1<\beta<\frac{u}{\A},\quad 0<\A<u<\B.&\end{eqnarray*} The
interpretation and the origin of this coordinate change is shown
in \figref{fig:support}. We finally get
\newcommand{\skp}{\!\!\!\!}
\begin{eqnarray}\label{eq:muIntegral} &&\skp
\int f(\vec{r},\vec{v})\ud{}^3{\vec{r}}\ud{}^3{\vec{v}}\,
\propto\int_{\A}^{\B}\ud{u}\,\sqrt{u}\,\mu_u[f], \qquad
\mathrm{where}\\  &&\skp\mu_u[f]=\!\! \int\limits_{u/\B}^{1}\!\!
\frac{\ud{\alpha}}{\sqrt{1-\alpha}} \!
\int\limits_{1}^{u/\A}\!\!\!\frac{\ud{\beta}}{\sqrt{\beta-1}}
\frac{\beta-\alpha}{\br{\alpha+\beta}^{5/2}}f\!\br{
\frac{\beta-\alpha}{\beta+\alpha},\frac{1}{u}\frac{\alpha\beta}{
\alpha+\beta}\!}.\nonumber\end{eqnarray} Now, given a $f(e,\e)$,
all expectation values can be determined in principle. In
calculating them, the following expressions are useful:
\begin{eqnarray} \label{g_vr_dispersion}
\!\frac{rv_r^2}{2\,GM}=\!\frac{\br{1-\alpha}\br{\beta-1}}{\alpha+\beta},\quad
\frac{rv_{\phi}^2}{GM}=\!\frac{2}{\alpha+\beta},\quad
e=\!\frac{\beta-\alpha}{\beta+\alpha}.\end{eqnarray} For the
purpose of further applications, the averages over concentric thin
spherical shells are important. The mean value of a function $g$
defined on a spherical shell of radius $r$, and consequently, the
average over all spherical shells can be calculated from
\begin{equation}\label{expectation_value}
\disp{g}_r=\frac{\mu_u[fg]}{\mu_u[f]}\quad \mathrm{and}\quad
\disp{g}=\frac{\int\ud{u}\sqrt{u}\,\disp{g}_u\mu_u[f]}{\int\ud{u}\sqrt{u}
\,\mu_u[f]},\end{equation} respectively. The expression for
$\disp{g}_r$ is quite analogous to a conditional probability `A
provided that B' for events $A$ and $B$.

\subsubsection{\label{sec:polyconstr}An orthogonal decomposition of
the phase space function} 
\begin{figure}
\centering
\includegraphics[height=
0.65\textheight
]{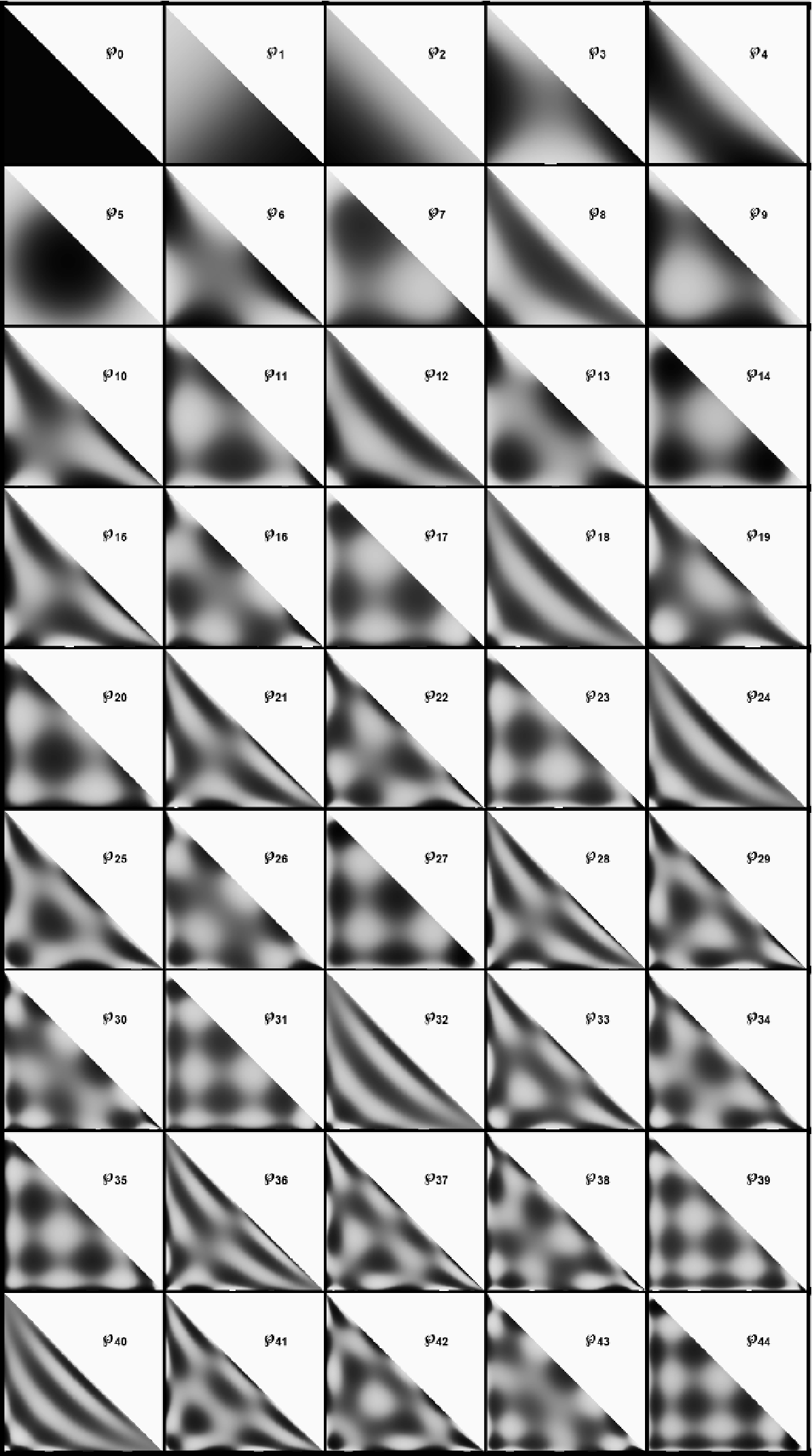}\\
\caption{\label{fig:TriangPolys} Contour maps (with a certain
automatic shading function) of $45$ initial polynomials from a
basis of polynomials orthonormal with respect to a standard scalar
product on the unit simplex $\xi+\eta-1<0$, $0<\xi<1$, $0<\eta<1$,
constructed using the procedure of \secref{sec:polyconstr}}
\end{figure}

The problem of finding a \pdf,
$f(e,\e)$,  on a triangular domain (\figref{fig:support}) can be
reduced to finding a series expansion of an auxiliary function
$h(\xi(e,\e),\eta(e,\e))$ in a basis of polynomials orthonormal on
a simplex in the plane $\xi,\eta$ (we assume that $f\equiv h^2$,
then $f$ is non-negative; the transformation  functions
$\xi(e,\e),\eta(e,\e)$ still need to be specified). To this end we
apply the Gram-Schmidt method with a scalar product
$g(u,v)=2\int_{0}^{1}\ud{\xi}\int_{0}^{1-\xi}\ud{\eta}\,u(\xi,\eta)
v(\xi,\eta)$ on the unit simplex $0<\xi<1$, $0<\eta<1$,
$\xi+\eta<1$. First, we define two families of polynomials of
degree that does not exceed a given $d>0$, namely, $A_{n,k}$:
$\eta^k\xi^{n-k}-\xi^k\eta^{n-k}$ and $S_{n,k}$:
$\eta^k\xi^{n-k}+\xi^k\eta^{n-k}$, with $0\leq n\leq d$, $0\leq
k\leq n$. By symmetry, any $A$ is orthogonal to any $S$. Then, we
sort polynomials $S$, to get a sequence with a nondecreasing
degree and take their union, thereby obtaining a reduced sequence
$S'$. We transform $S'$ to another sequence by the consecutive
projections
$s'_m=S'_m-\sum_{i=1}^{m-1}\frac{g(S'_m,S'_i)}{g(S'_i,S'_i)}S'_i$
and normalize $s'_m\to s_m=\br{g(s'_m,s'_m)}^{-1/2} {s'_m}$. We
repeat the same procedure for nonzero polynomials $A$ to obtain a
sequence $a_{m'}$. Finally, we take the union of sequences $s_m$
and $a_{m'}$, and sort with respect to the increasing degree,
obtaining the required basic polynomials $\p_j(\xi,\eta)$,
$j=0,1,\dots,(d+1)(d+2)/2$. The initial $45$ polynomials obtained
this way are shown in \figref{fig:TriangPolys}.
Finally, we make a coordinate change
$\xi=\frac{u_b+u_a}{u_b-u_a}e$,
$\eta=\frac{u_a}{u_b-u_a}\br{2u_b\e-e-1}$, thereby mapping the
simplex to the triangular domain \textbf{ABC} in
\figref{fig:support}. This way, the task of finding an orthonormal
basis, of $D_d=(d+1)(d+2)/2$ polynomials
$\q_j(e,\e)=\p_j\br{\frac{u_b+u_a}{u_b-u_a}e,\frac{u_a}{u_b-u_a}\br{2u_b\e-e-1}}$
of degree not exceeding $d$ in $e$ and $\e$, has been completed.

\subsubsection{\label{sec:finding} Finding the phase space function and the derived
quantities}

The theoretical \rvd can be compared with the quantity
$\disp{rv_r^2}$ determined based on measurements for a large
enough sample of points $r_i$. Given $d$ and mass $M$, this
comparison enables us to find a set of optimal coefficients in the
following expansion of the phase space function $f(e,\e)$:
\begin{equation}\label{eq:expansion}f(e,\e)=h^2(e,\e),\qquad h(e,\e)\approx
\sum_{k=1}^{D_d}h_k \q_k(e,\e),
\end{equation} by
minimizing the following mismatch function:\footnote{To measure
the accuracy of a fit $Y(x)$ to data points $(x_i,y_i)$, we use a
dimensionless number $\delta$ such that
$\delta^2=\frac{1}{N\overline{y}^2}\sum_i\br{Y(x_i)-y_i}^2$, where
$\overline{y}=\frac{1}{N}\sum_{i}|y_i|$. The definition of
$\delta_M$ agrees with that of $\delta$.}
\begin{equation}\label{error_function}
\delta_M=\sqrt{\frac{\frac{1}{N}\sum_i
\left(\bar{\sigma}(u_i)-\avg{\frac{r_iv_r^2(r_i)}{GM}} \right)^2}
{\br{\frac{1}{N}\sum_i\avg{\frac{r_iv_r^2(r_i)}{GM}}}^2}},\qquad
u_i=r_i/R,
\end{equation}
$N=\sum_i1$ (it suffices that $N$ be several times greater than
$D_d$), where
\begin{equation}\label{eq:SigmaExpans}\bar{\sigma}(u)=\frac{
\sum_{k,l}^{D_d}h_k
h_l\,\mu_u\sq{\sigma\p_k\p_l}}{\sum_{k,l}^{D_d}h_k
h_l\,\mu_u\sq{\p_k\p_l}},\quad
\sigma=\frac{2(1-\alpha)(\beta-1)}{\alpha+\beta},\end{equation}
with $\frac{\beta-\alpha}{\beta+\alpha}$ and
$\frac{1}{u}\frac{\alpha\beta}{\alpha+\beta}$ substituted  in
$\p${}s for $e$ and $\e$, respectively. The integrals
$\mu_u\sq{\p_k\p_l}$ and $\mu_u\sq{\sigma\p_k\p_l}$ defining
functions of a single argument $u$ leads to integrals of the
general form:
$$\frac{1}{u^s}\!\!\!\int\limits_{u/u_b}^1\!\!\!\ud{\alpha}
\!\!\!\int\limits_{1}^{u/u_a}\!\!\!\ud{\beta}\,
\frac{\alpha^p\beta^{q}}{(\alpha+\beta)^r\,
\sqrt{(1-\alpha)(\beta-1)(\alpha+\beta)}},$$ with integers
$p,q,r,s$. The analytical form of this integral can be found
recursively for any set of integers $p,q,r$.\footnote{We have not
found any general formula for such an integral; however, having
written a code using a computer algebra system, we were able to
find the integrals in analytic form by recursions for any $p,q,r$.
The alternative way -- a numerical integration, would be
inefficient in the minimization procedure; we observed that either
the errors accumulated in an uncontrolled fashion, resulting in a
high noise level in $\delta_M$, or the computation was too time
consuming when the precision was increased to keep the errors
within acceptable bounds.}

Since $\delta_M$ is homogenous with degree $1$ in the variables
$h_k$, it only depends on points on a $(D_d-1)$-dimensional unit
sphere. To find the $\delta_M$, one starts with a random point on
that sphere (with the help of a generator uniform on that sphere),
and then the minimization procedure is run. There can be many such
minima, or a minimum can be degenerated. Having found $h_k{}$'s
with this method, secondary quantities other than
$\bar{\sigma}(u)$ can be computed, using definitions like
\eqref{g_vr_dispersion}. They lead to expressions $\bar{e}(u)$,
$\bar{\beta}(u)$, etc., of the same general form as that of
$\bar{\sigma}(u)$ in \eqref{eq:SigmaExpans} (with quantities of
the kind as in \eqref{g_vr_dispersion} substituted for $\sigma$ in
$\mu_u\sq{\sigma\p_k\p_l}$), that is, a quotient of two quadratic
forms in $D_d$ variables $h_i$.

\section{The results}\label{sec:results}

\begin{figure}
\hspace{-0.015\textwidth}
\includegraphics[width=0.525\textwidth]{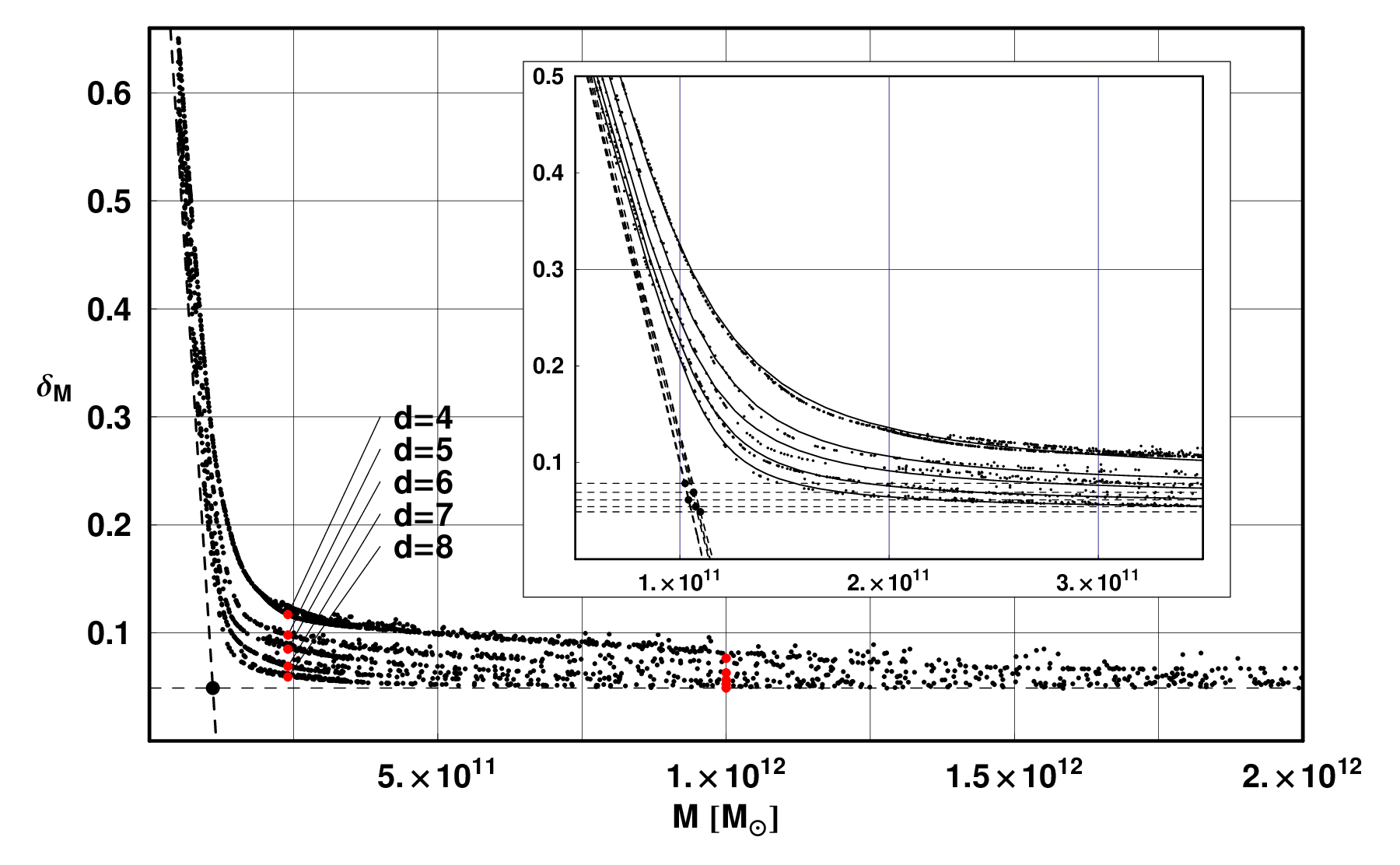}\\
\caption{\label{fig:ChiMass} A $\chi^2$-like-test defined by
$\delta_M$ for best-fit curves to the \rvd profile of sample \tI
(see \figref{fig:RVDprofiles}, bottom panel) shown as a function
of the central mass $M$. The phase space support used to obtain
these results was fixed by $u_a=18\kpc$ and $u_b=300\kpc$, and the
fitting region was limited to $r>22\kpc$. One can distinguish $5$
accumulation curves that correspond to various degrees $d$ of
polynomials used to approximate the phase distribution function
(here, $d=4,5,6,7,8$). These curves were compared with
best-fitting hyperbolas in the inset figure.  Abscissae of
crossing points of the corresponding asymptotes suggest the
presence of a lower bound for mass in the limit of large $d$. The
red dots represent minimum values of $\delta_M$ found at
$M=2.4\times10^{11}\msun$ and, to compare with, at
$M=1.0\times10^{12}\msun$ (for $d=8$ also points with
$u_b=240\kpc$ were included).}
\end{figure}

\begin{figure}
\hspace{0.008\textwidth}
\begin{tabular}{@{}c@{}}
\hspace{-0.012\textwidth}
\includegraphics[width=0.486\textwidth]{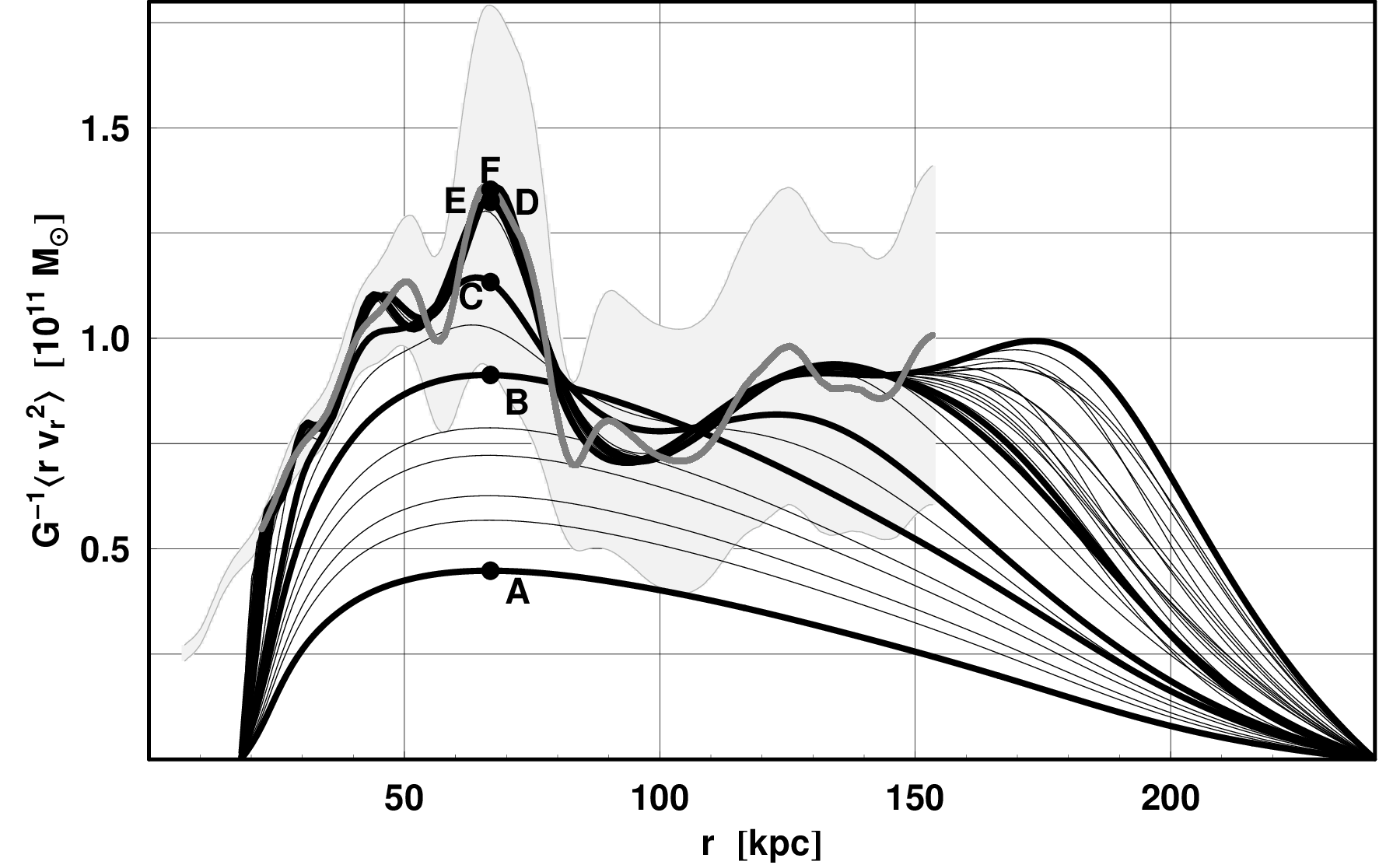}\\
\hspace{-0.03\textwidth}
\includegraphics[width=0.500\textwidth]{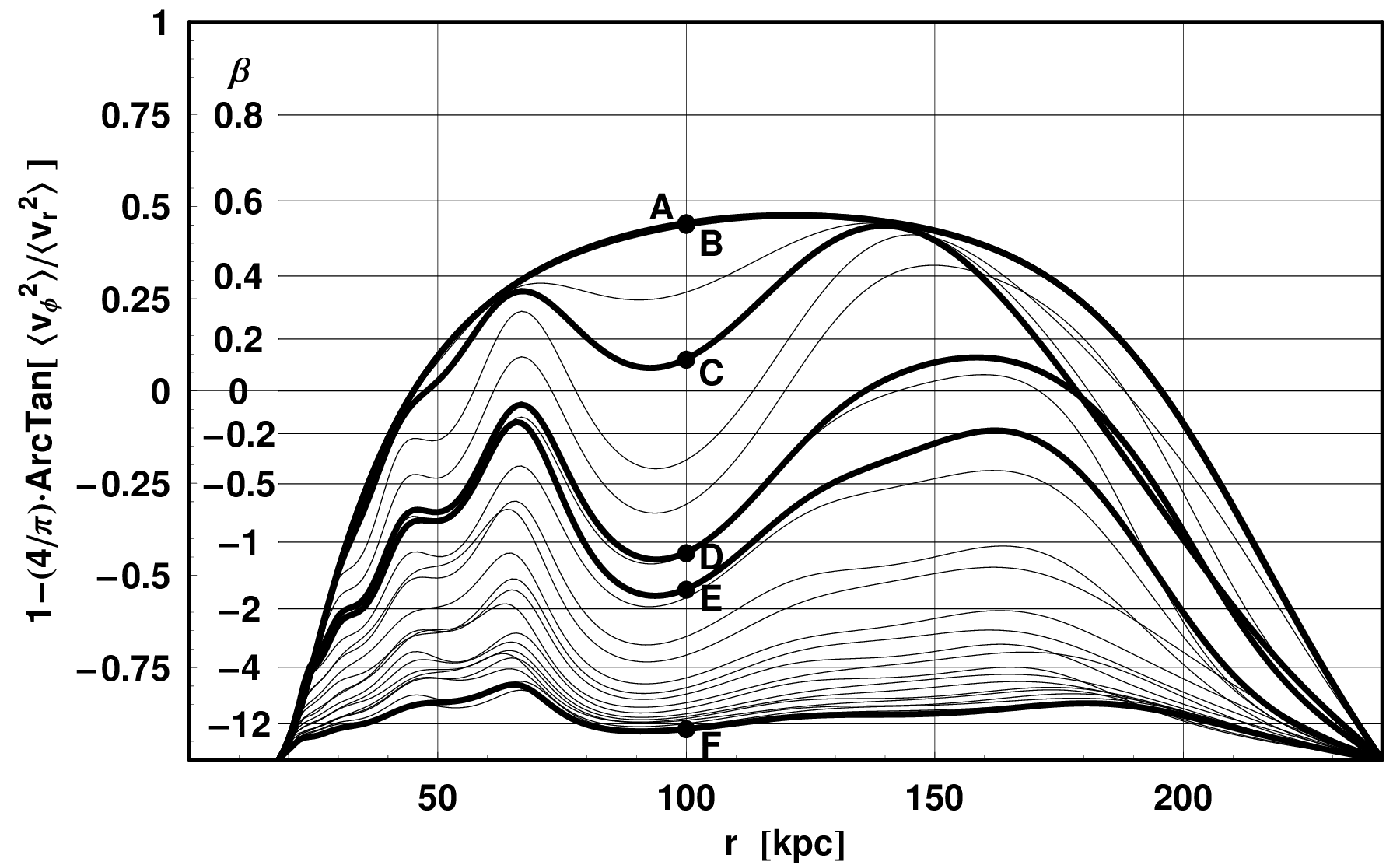}\\
\hspace{-0.013\textwidth}
\includegraphics[width=0.500\textwidth]{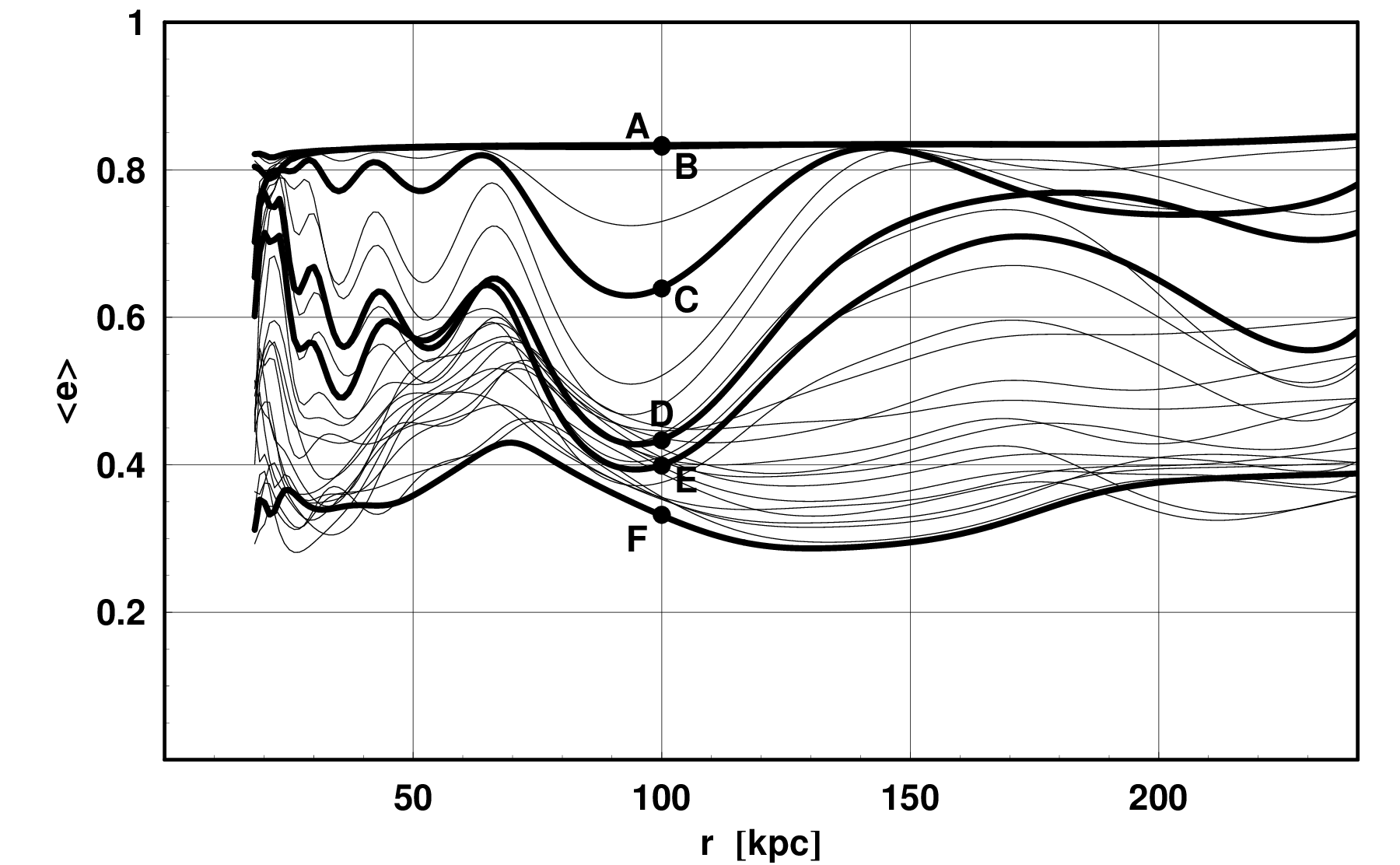}\\
\end{tabular}
\caption{\label{fig.variousmasses}
 \textit{[Top]} --- example best-fit curves to the
 \rvd profile
 of sample \tI for
masses ranging from $0.5\times10^{11}\msun$ to
$10\times10^{11}\msun$ (A: $0.5\times10^{11}\msun$, B:
$1.02\times10^{11}\msun$, C: $1.3\times10^{11}\msun$, D:
$2.09\times10^{11}\msun$, E: $2.4\times10^{11}\msun$, F:
$10.2\times10^{11}\msun$), assuming $d=8$ and $u_a=18\kpc$,
$u_b=240\kpc$ ($\Rightarrow$ $0<e<0.83$). The light gray region is
the uncertainty in the \rvd profile corresponding to vertical bars
in \figref{fig:RVDprofiles}, \textit{[middle]} the corresponding
(symmetrized) anisotropy parameter, and \textit{[bottom]} the mean
eccentricity.}
\end{figure}

Using the minimization method of \secref{sec:finding}, we started
by generating several best fit curves to the \rvd profile of the
\tI sample of tracers and assumed various central masses $M$,
ranging from $0.5\times10^{11}\msun$ to
$50\times10^{11}\msun$.\footnote{Most of the values in this range
of masses are physically implausible and were only used for
illustration.} This allowed us to compare the quality of best fits
measured by the mismatch function $\delta_M$ defined in
\eqref{error_function} and to see how it changes with the number
$D_d$ of polynomials used in \eqref{eq:expansion} to approximate
the \pdf. The result of this comparison is presented in
\figref{fig:ChiMass},
where $\delta_M$  is shown versus the central mass $M$ for various
$d$. The example best-fit curves to the \rvd profile (assuming
$d=8$) were compared in \figref{fig.variousmasses}.
Lower panels in \figref{fig.variousmasses} show the corresponding
secondary quantities. As expected, they can change significantly
when \rvd best-fit curves are not changed at all. This illustrates
our earlier theoretical expectation that $M$ cannot be determined
based on the measurements of the radial motions alone. Also, due
to the flattening of the $\delta_M$ curves in \figref{fig:ChiMass}
for high masses, the optimum mass, chosen to correspond to the
lowest $\delta_M$ fit (a global minimum), would have unacceptably
large uncertainty (as would the secondary quantities that are
largely $M$-dependent). Thus, the criterion of mass must be
different. Nevertheless, one can estimate a lower bound for $M$.

As can be seen in \figref{fig:ChiMass}, for high enough $M$,
$\delta_M$ gets reduced when $d$ is increased and appears to tend
to some small nonzero limit, and the same for all $M$ when $d$ is
large enough. For lower $M$, $\delta_M$ decreases faster with
increasing $d$ than previously, but for $d$ large enough the
attained limit seems the same as for larger $M$. For even lower
masses, below $2\times10^{11}\msun$, the mismatch function
$\delta_{M}$ grows rapidly. With low enough $M$, therefore, it is
not possible to obtain a satisfactory best fit curve. The diagram
in \figref{fig:ChiMass} suggests that there is a sharp lower bound
for $M$ in the limit of large $d$. The presence of such a lower
bound can be inferred from the observation that for a given
$\delta$, $d$ can be increased so that $M$ is reduced
significantly to a value $M-\Delta M$ with the same quality
criterion $\delta_M=\delta_{M-\Delta M}=\delta$ fulfilled. For
example, as seen in \figref{fig:ChiMass}, a model with $d=6$ and a
high mass $M=10\times10^{11}\msun$ (a value that would be accepted
without hesitation in the dark-matter halo paradigm) passes the
same $\chi^2$-like-test as another model with $d=8$ and more than
four times lower mass $M=2.4\times10^{11}\msun$. With larger $d$,
the same level of $\delta$ would be attained even for a lower mass
$M$. The crossing point of the asymptotes in \figref{fig:ChiMass}
obtained based on the shown data, determines the approximate value
for the expected minimal $\delta$ possible in the theoretical
limit $d\to\infty$ (the saturation value), and the corresponding
minimal mass -- the lower bound for $M$.

Overall, the minimum value for $\delta_M$, at any $M$ from the
right-hand neighborhood of the lower bound for masses, appears to
saturate in the limit of large $d$ and remains invariable in a
wide range of masses. (Already for $d=8$ the $\delta_M$ is nearly
saturated for larger masses, therefore -- except for leading to
insignificant reduction in $\delta_M$ -- considering $d>8$
introduces nothing new to our qualitative analysis.) The presence
of the nonzero saturation value for $\delta_M$ indicates that the
particular shape of the \rvd curve cannot be exactly accounted for
by the Keplerian ensemble model, irrespective of the assumed mass
(but perfection is not what is expected from a model). The really
important lesson we can learn from the model is that for all
masses greater than some limit, the saturation value for
$\delta_M$ is low and comparable. It means that the best-fit
curves to the \rvd profile are equally good, independent of mass.
Putting this differently, the application of the $\chi^2$-like
criterion defined by $\delta_M$ cannot distinguish between the
admissible masses, since one cannot tell any difference between
the corresponding best fits: the mass could be low (but bounded
from below), as well as arbitrarily high. The \rvd profile alone
is thus insufficient for sharply determining the real mass,
although it suffices for estimating the lower bound for admissible
masses. Some other observables or conditions imposed on them must
be taken into account to constrain the \mw mass. However, these
constraints should be imposed with due care in order not to
overestimate the \mw mass unnecessarily.

\begin{figure*}
\centering
\begin{tabular}{|@{}c@{}c@{}|}
\hline {}&{}\\
\includegraphics[width=0.50\textwidth]{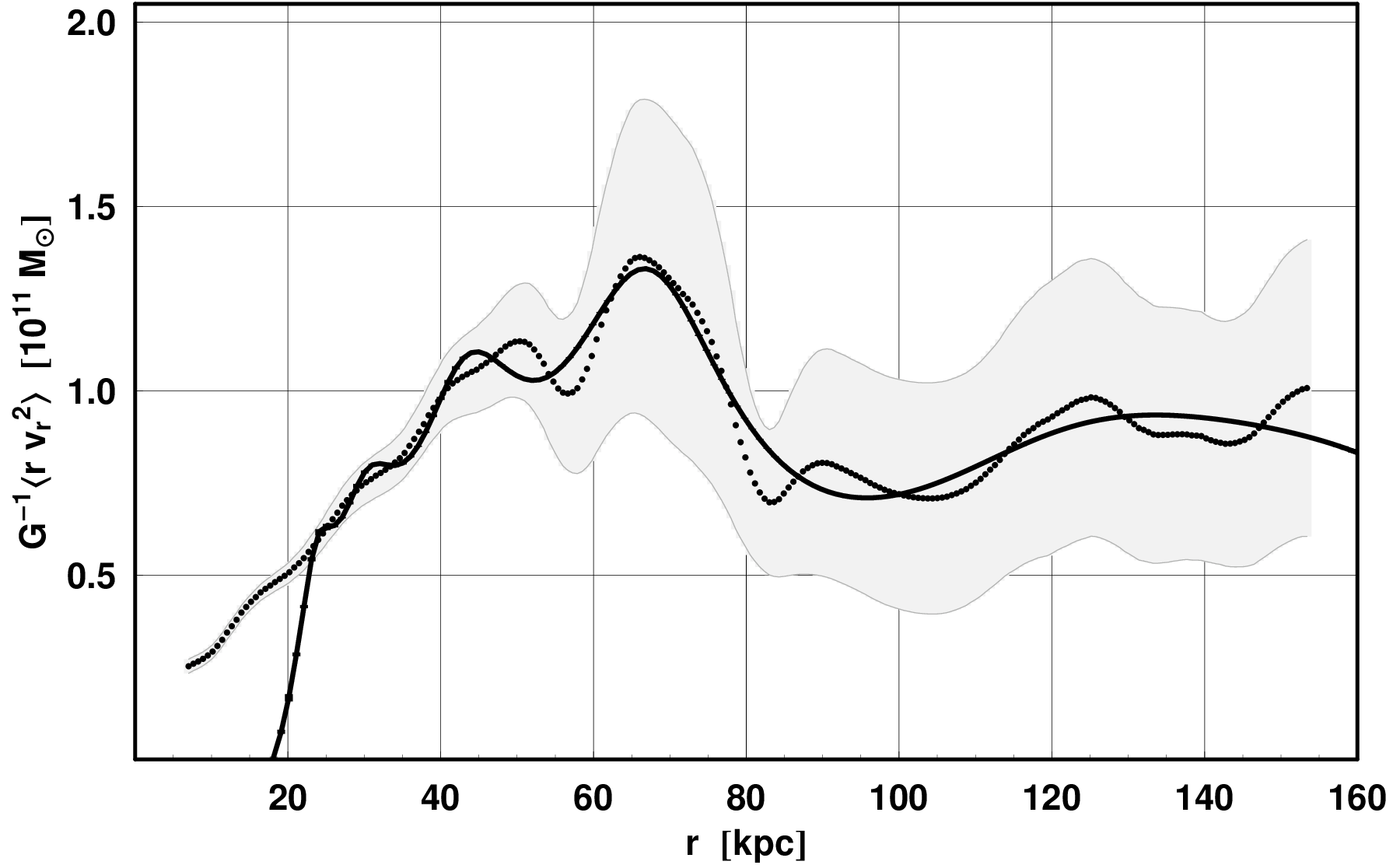}&
\includegraphics[width=0.50\textwidth]{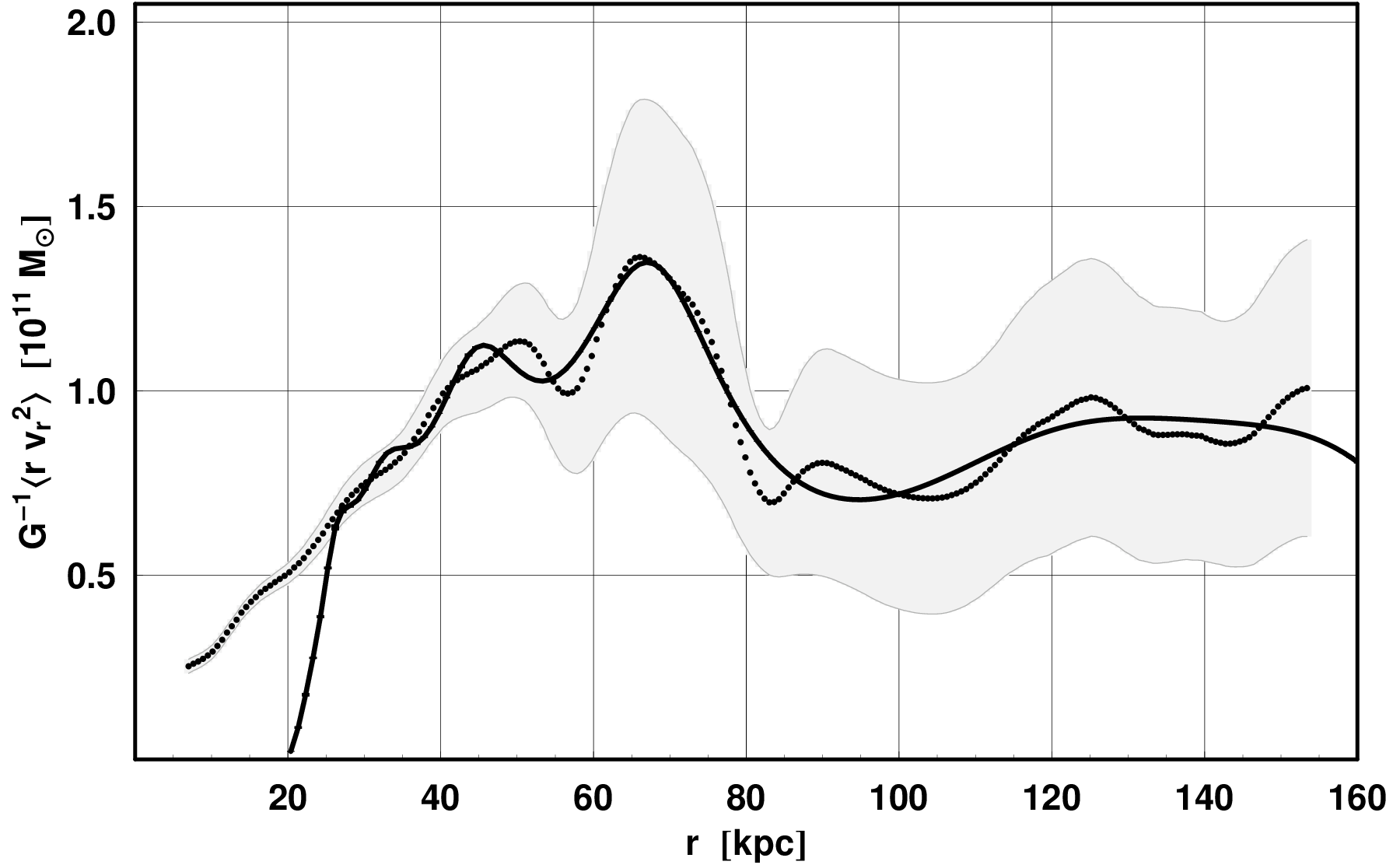}\\
\tI; $\quad r_a=18\kpc$, $r_b=240\kpc$, $r_s=22\kpc$ & \tI; $\quad
r_a=20\kpc$, $r_b=220\kpc$,
 $r_s=25\kpc$ \\
 \hline  \hline {}&{}\\
\includegraphics[width=0.50\textwidth]{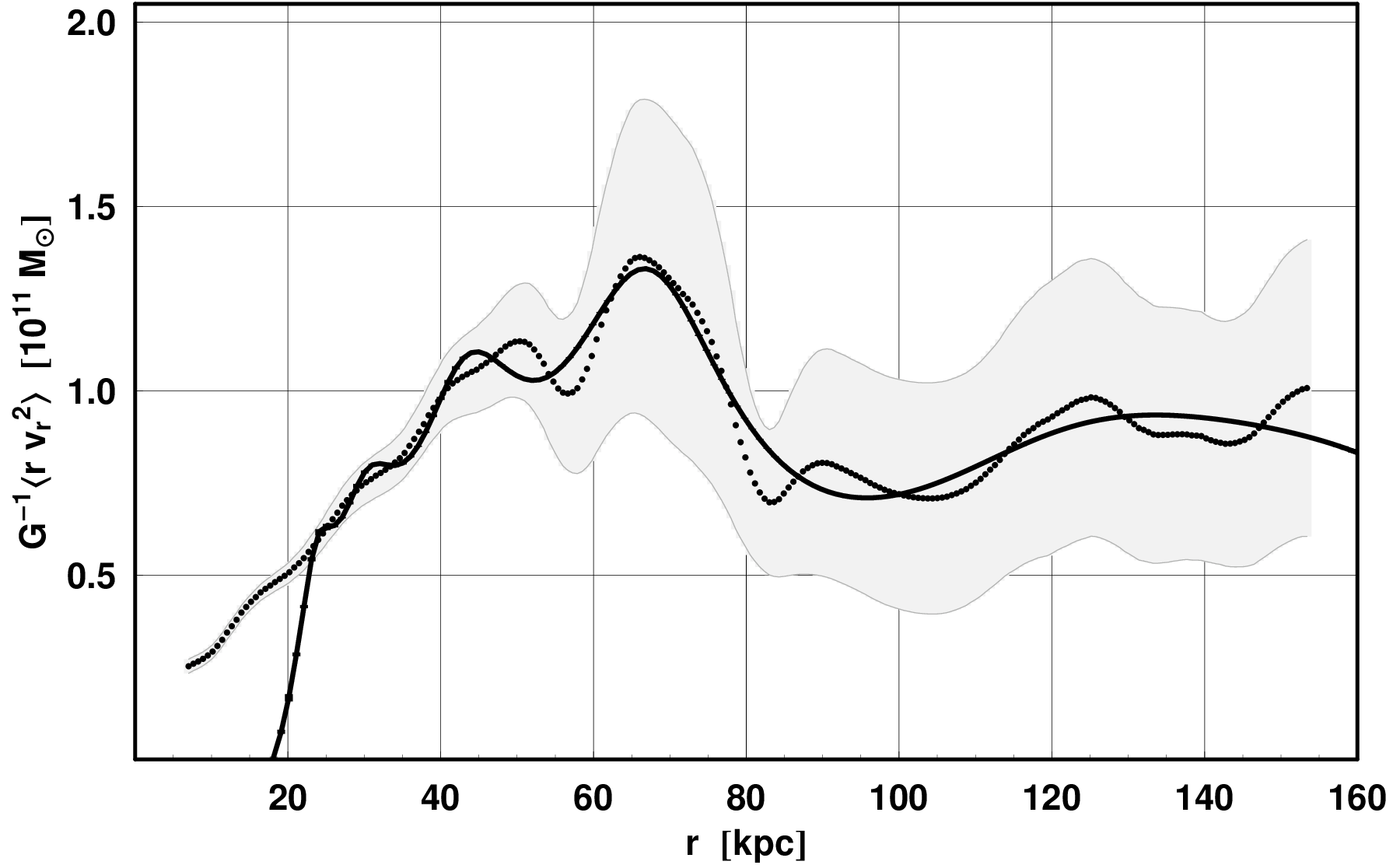}&
 \includegraphics[width=0.50\textwidth]{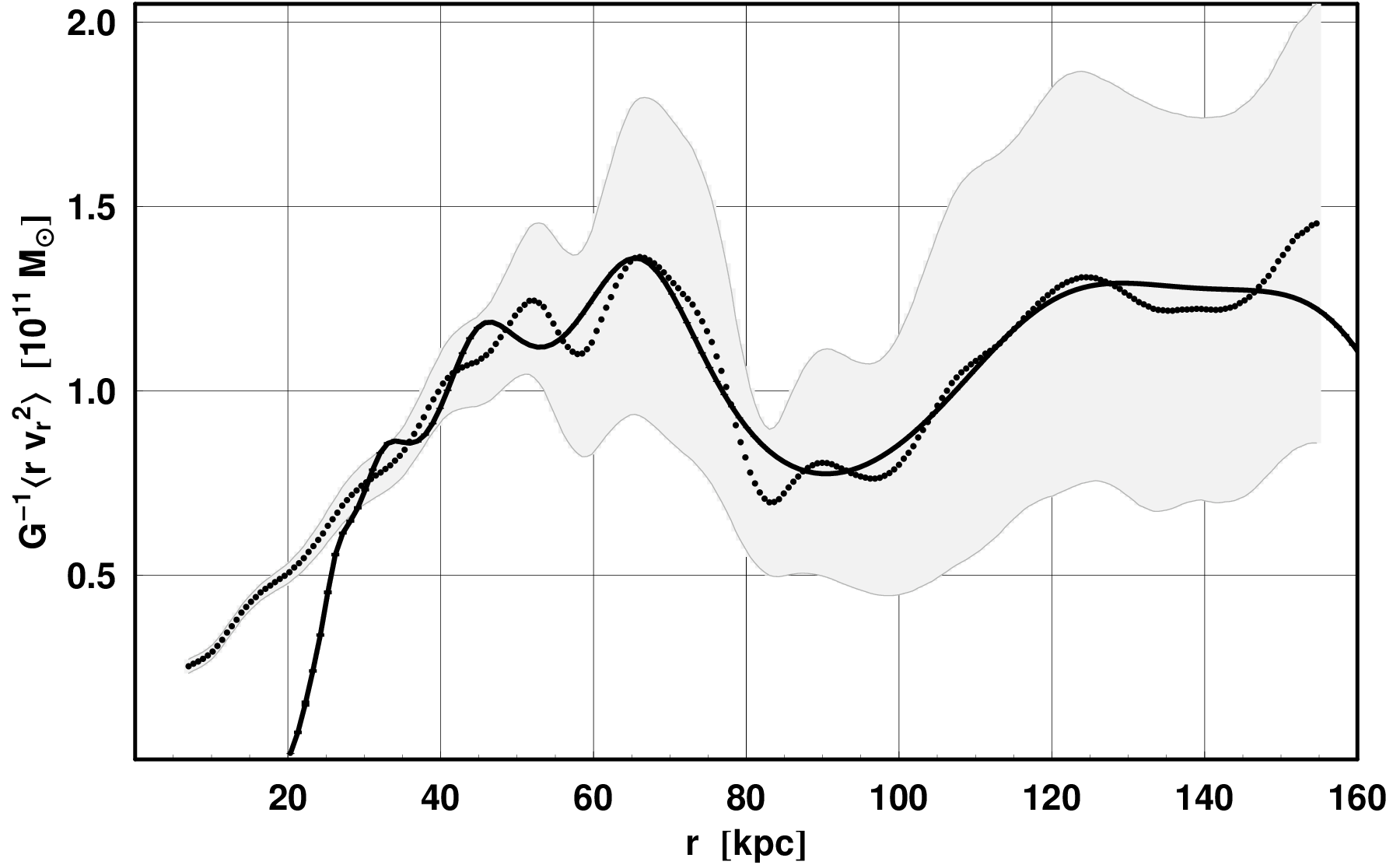}\\
\tII; $\quad r_a=18\kpc$, $r_b=240\kpc$, $r_s=22\kpc$ & \tII;
$\quad r_a=20\kpc$, $r_b=250\kpc$,
 $r_s=25\kpc$\\
\hline
\end{tabular}
\caption{\label{fig:RVD} \textit{[thick lines]} -- the mean of
best-fit curves
 to radial velocity profiles (\rvd);
\textit{[dotted lines]} -- the measured \rvd profiles of samples
\tI and \tII, obtained with the help of the minimization procedure
of \secref{sec:finding} assuming $d=8$ and
$M=2.4\times10^{11}\msun$. (The curve's width equals twice the
standard deviation from the mean.) The light gray region is the
uncertainty in the \rvd profile corresponding to vertical bars in
\figref{fig:RVDprofiles}. Here, $r_s$ is the left bound for the
fitting region, while $r_a$ and $r_b$ determine the triangular
support \textbf{ABC} in \figref{fig:support}. }
\end{figure*}

We have seen that $M$ cannot be determined with the help of the
best-fit criterion. Instead, one can assume a certain $M$ and see
if the resulting secondary quantities are plausible. As an
example, we take the reference mass $M=2.4\times10^{11}\msun$,
which was inferred in the past from the Galaxy rotation inside
$20\kpc$ \citep{1992AJ....103.1552M}. This mass value also
overlaps with the median value $2.4^{+1.3}_{-0.7}\times
10^{11}\msun$, found in point potential with isotropic velocities
\citep{1987ApJ...320..493L} for a sample of ten satellites in the
Galactic halo at distances $50-140\kpc$ (not including \leoI),
which was deemed the most reliable one among other samples
considered by these authors. Assuming the reference mass, we
performed a large number of minimizations, starting from various
initial points chosen randomly on the $(D_d-1)$-dimensional unit
sphere of expansion coefficients $h_k{}$. Normally, one would
expect convergence to a unique minimum (or several isolated
minima), regradless of the starting point. Instead, we found a
submanifold of minima on that sphere, giving rise to a number of
best-fit curves through the measured \rvd, as shown in
\figref{fig:RVD} (upper panels). Although the sets of expansion
parameters (thus also the corresponding \pdf{}s) are different,
the corresponding best-fit curves to the same profile of the
radial velocity dispersion are almost indistinguishable.
\begin{figure*}
\centering
\begin{tabular}{@{}r@{}}
\begin{tabular}{@{}r@{}r@{}r@{}}
\includegraphics[width=0.5\textwidth]{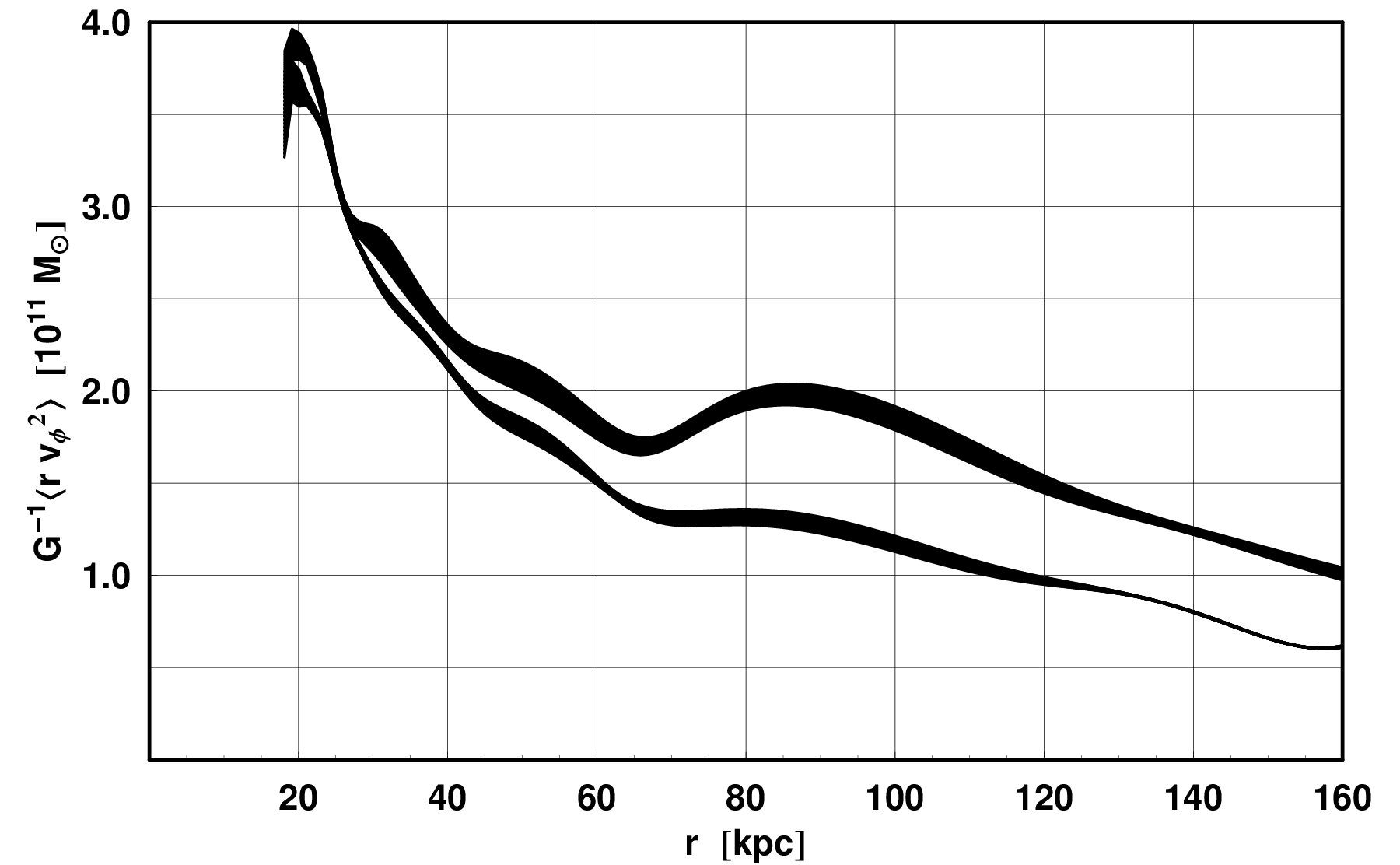}&
\includegraphics[width=0.5\textwidth]{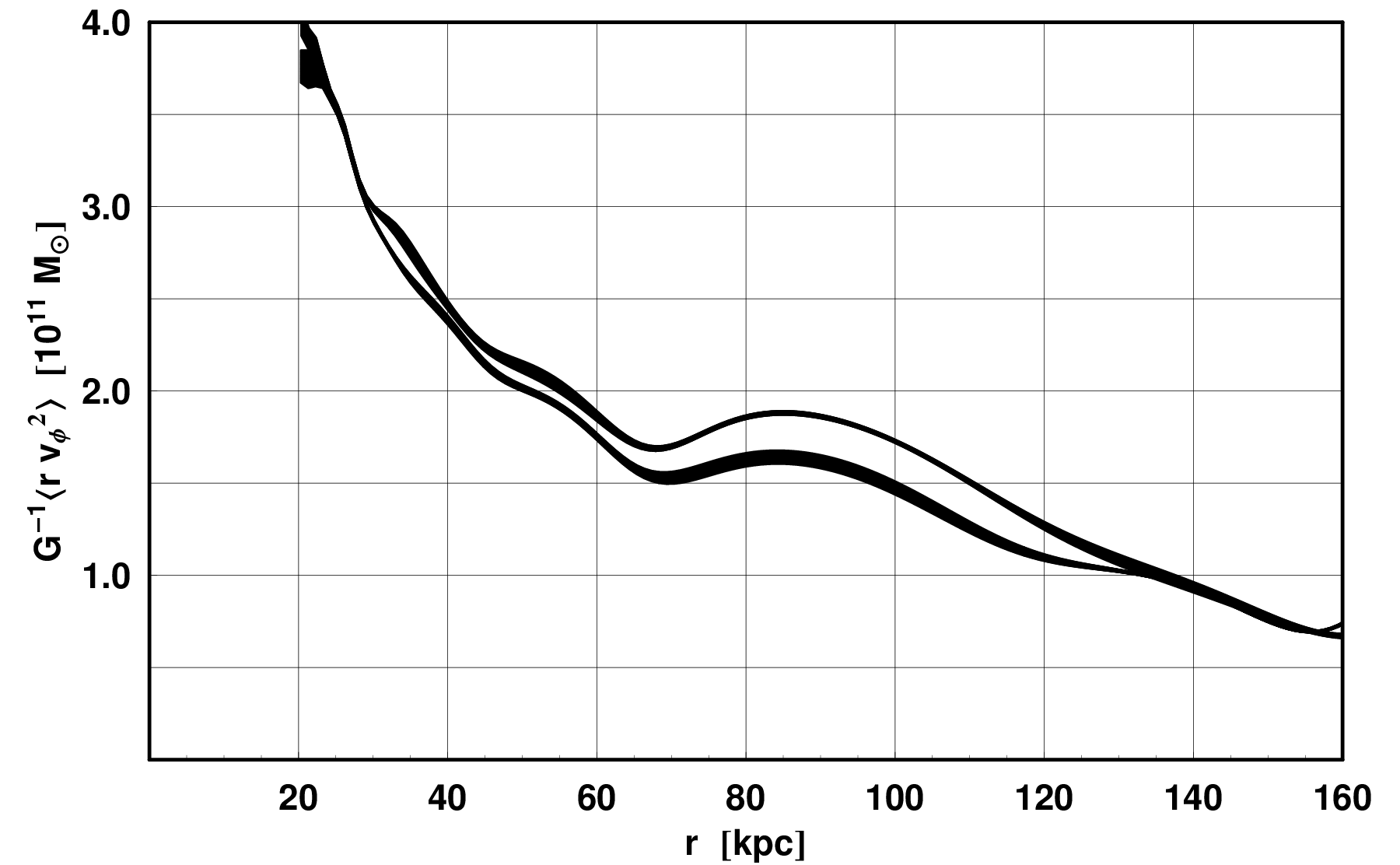}&
\hspace{-0.0\textwidth}\\
\end{tabular}\\
\begin{tabular}{@{}r@{}r@{}r@{}}
\includegraphics[width=0.50\textwidth]{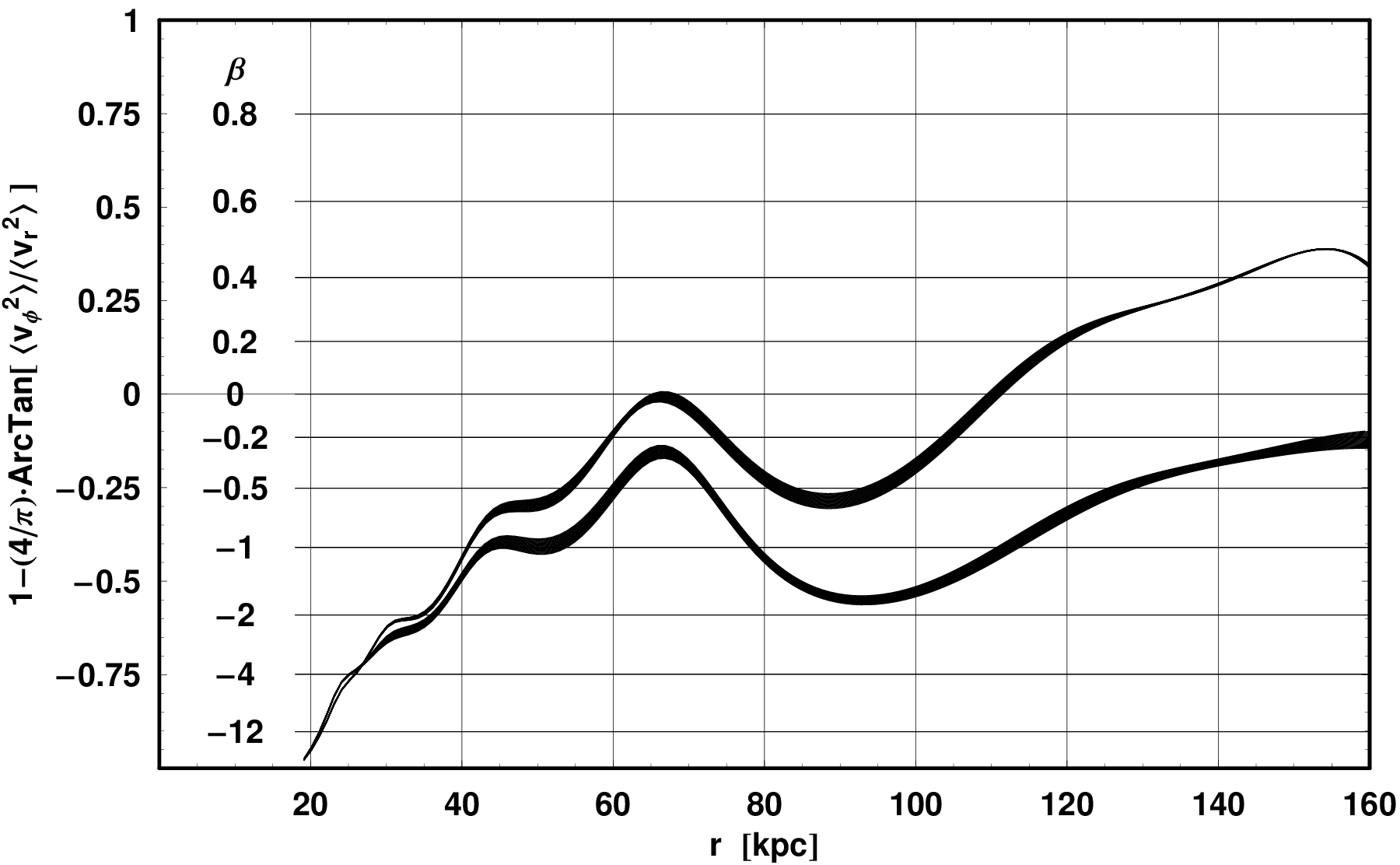}&
\includegraphics[width=0.50\textwidth]{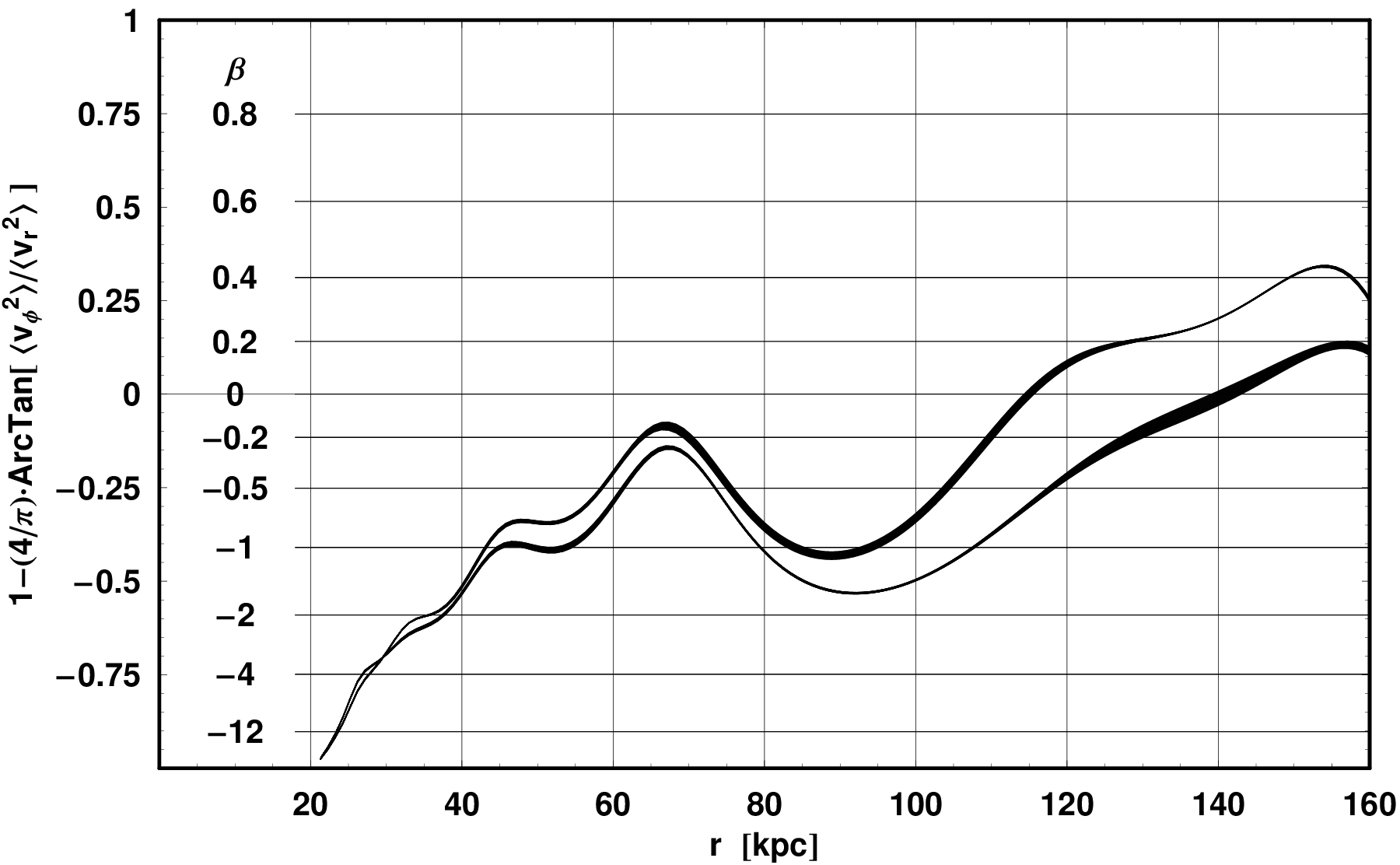}&
\hspace{+0.0075\textwidth}
\end{tabular}\\
\begin{tabular}{@{}r@{}r@{}r@{}}
\includegraphics[width=0.505\textwidth]{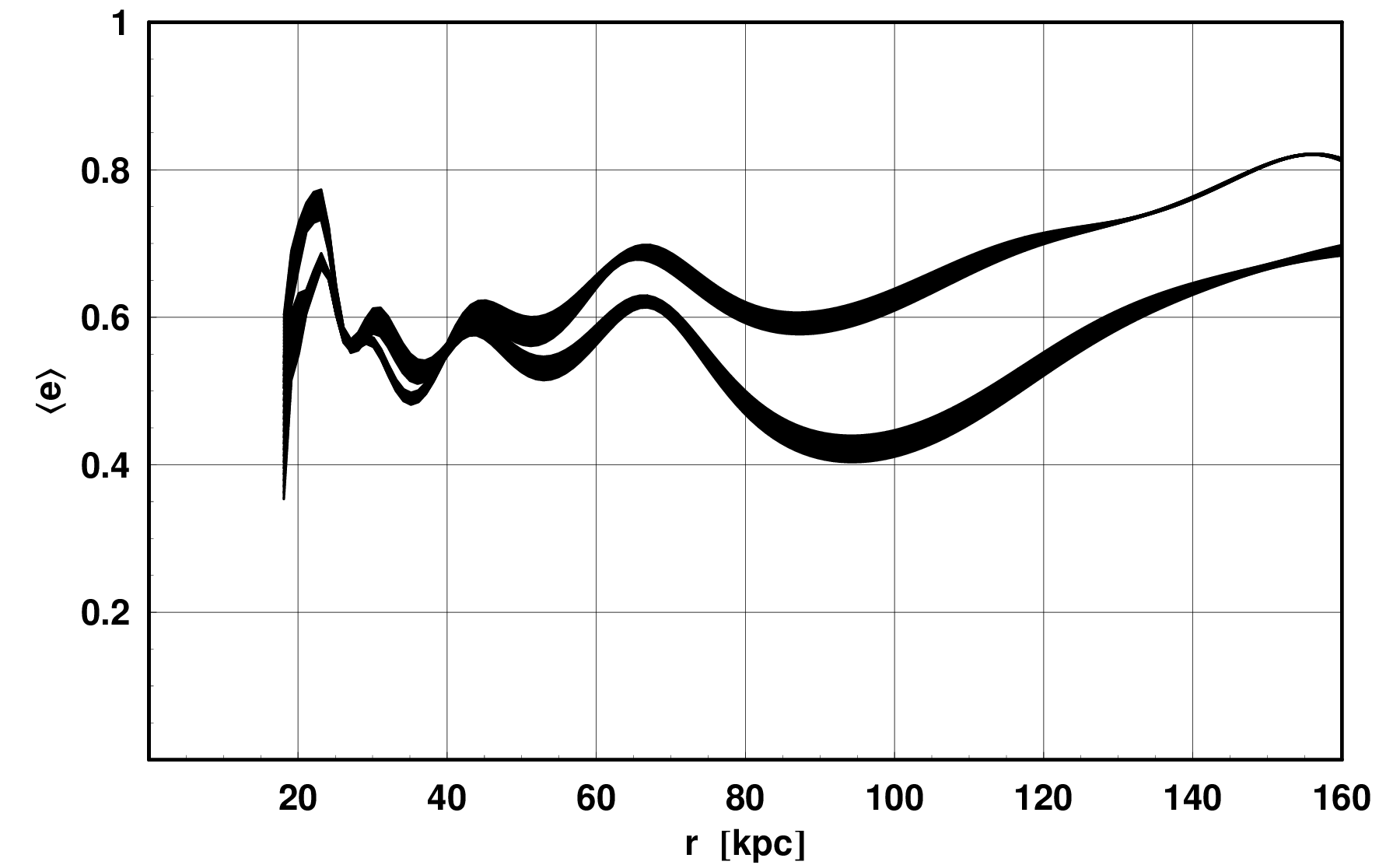}&
\includegraphics[width=0.505\textwidth]{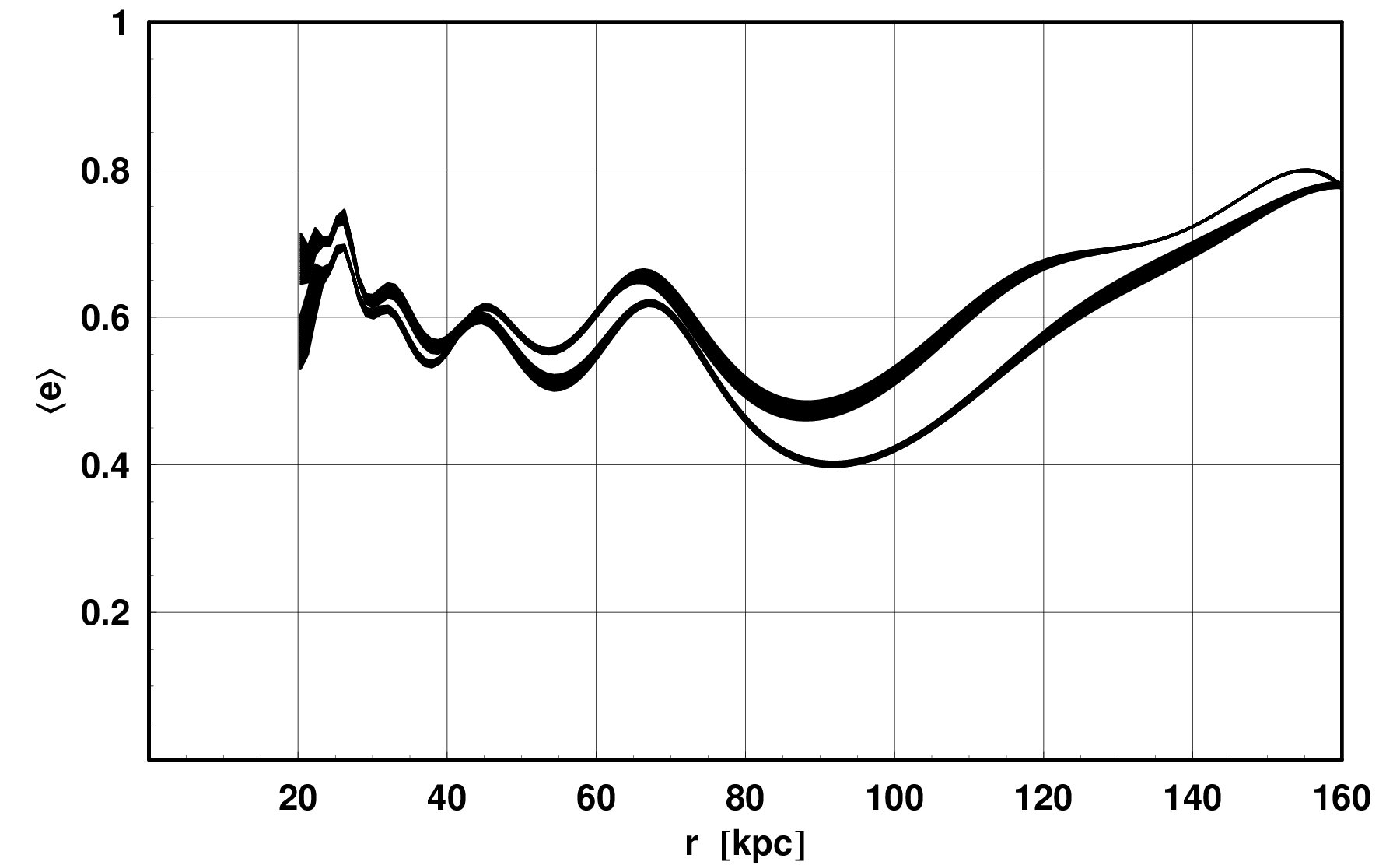}&
\hspace{+0.01\textwidth}\\
\end{tabular}\\
\begin{tabular}{@{}r@{}r@{}r@{}}
\hspace{-0.24\textwidth}\includegraphics[width=0.5\textwidth]{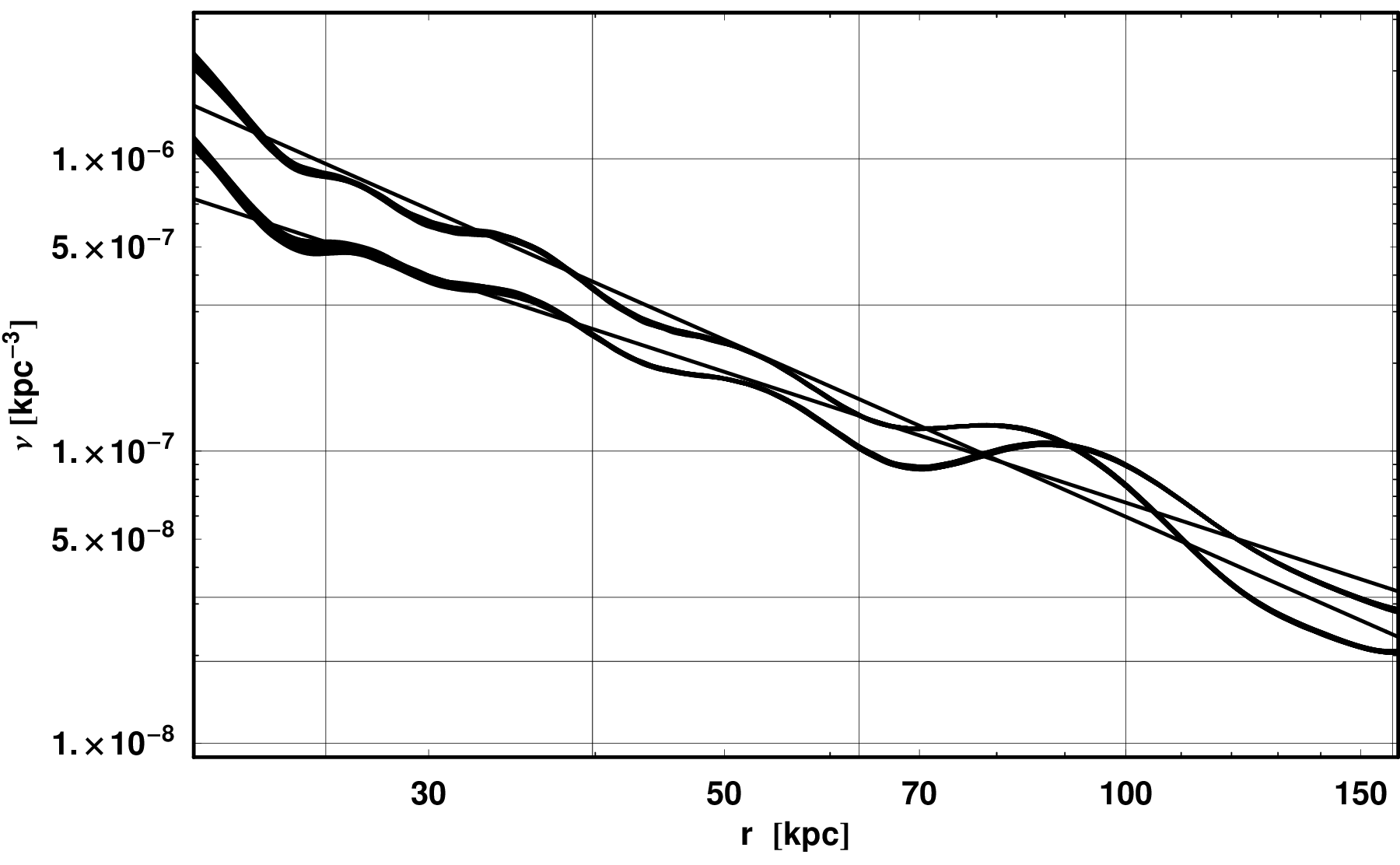}&
\includegraphics[width=0.5\textwidth]{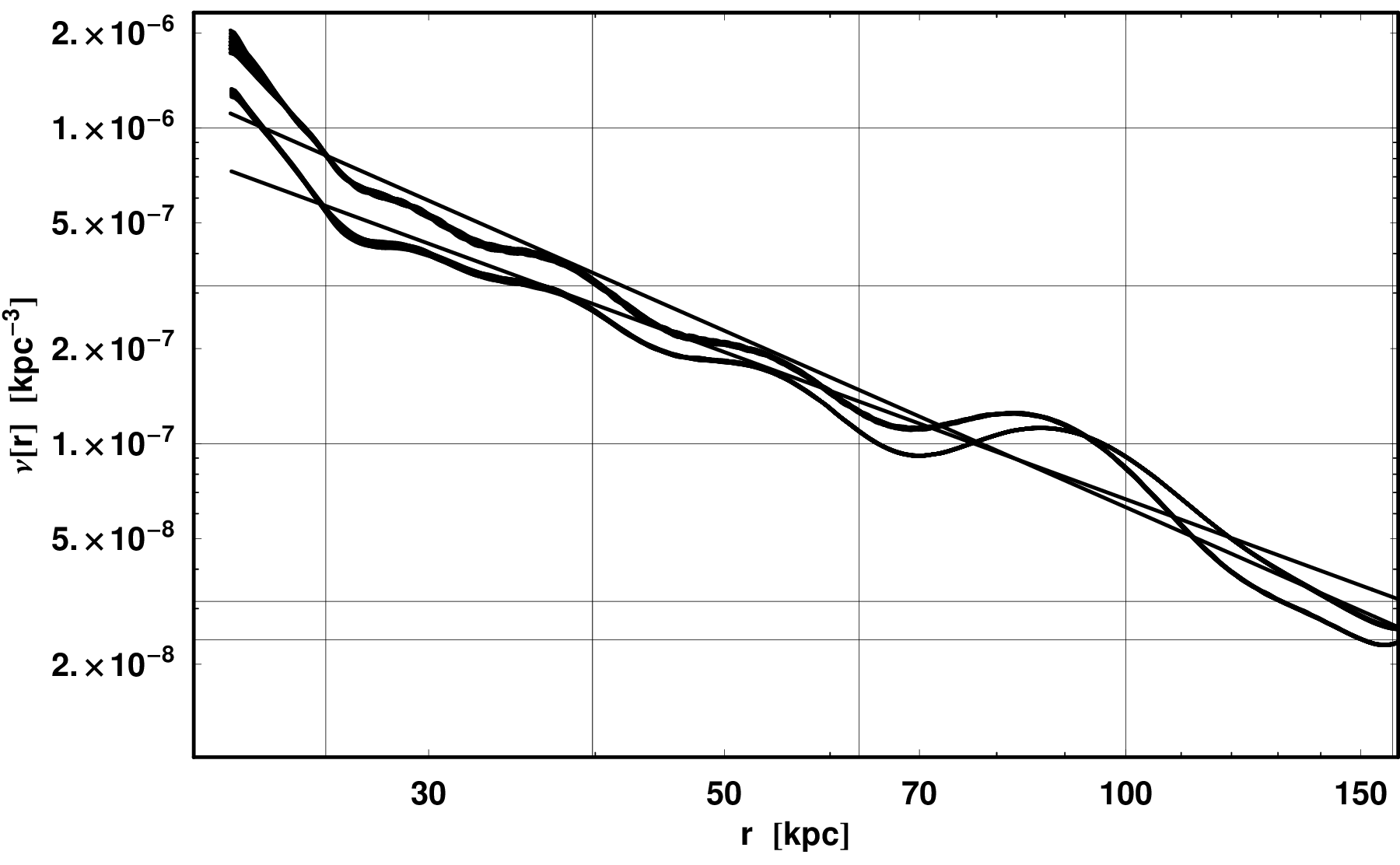}&
$\phantom{....}$\\
\end{tabular}\\
\end{tabular}
\caption{ \label{fig:Results} The mean value and spread of
secondary quantities for models that are within the uncertainty
range in \figref{fig:RVD}: dispersion of transversal velocity
$G^{-1}\avg{r v_{T}^2}$,  [$\avg{r v_{T}^2}\equiv\avg{r
v_{\phi}^2}=\avg{r v_{\theta}^2}$ by symmetry] \textit{[top]}; the
(symmetrized) anisotropy parameter
$1-\br{4/\pi}\mathrm{ArcTan}\br{\avg{v_{\phi}^2}/\avg{v_r^2}}$ (an
alternative to $\beta=1-\br{\avg{v_{\phi}^2}/\avg{v_r^2}}$)
\textit{[middle]}; and $\avg{e}$ \textit{[bottom]}; and the
implied number density of objects in "log-log" scale (integrable
to $1$ over the shown interval) with a power law fits $\nu\sim
r^{-\gamma}$ with $\gamma$ $1.49$ and $2.01$ (left) and $1.55$ and
$1.86$ (right); all derived from the \rvd{}s shown in
\figref{fig:RVD}.}
\end{figure*}
This degeneracy was to be expected, based on the indeterminacy of
the Jeans problem discussed in the introduction. Choosing various
sets of $h_k$ is tantamount to considering various solutions of
Jeans equation with the same mass model. Hence, while a best-fit
curve is not changed when $h_k{}$'s are varied inside that
manifold, the corresponding secondary quantities may change. This
is indeed the case, which can be seen in \figref{fig:Results}
(lower panels), where the most important secondary quantities
corresponding to the best-fit curves are shown. Interestingly,
this change in the secondary quantities is not dramatic, since the
spread of secondary quantities about their mean values is quite
moderate.

Another source of degeneracy in the space of
parameters $\{h_k\}$
 comes from the uncertainty in the \rvd profile.
 When one makes some variation of
 the \rvd within the uncertainty bars and then performs the
 minimization procedure, the resulting expansion parameters
 $\{h_k\}$ will remain
 in some small neighborhood of the initial minimum.
 Therefore, there are many distribution functions
 $f(e,\epsilon)$  consistent with
 the \rvd observations within a given uncertainty limit (error bars).
 This shows that the observational constraints other than
 \rvd are needed
 to narrow down the range of admissible distribution functions.

\section{The summary and concluding remarks}

We conjectured that galactic masses may be overestimated by
restricting the variety of solutions available in the framework of
unconstrained Jeans modeling. These restrictions arise through
imposing some subsidiary conditions when finding solutions to
Jeans equations. The conditions concern mainly the velocity
anisotropy profile, the form of which is often assumed and, most
frequently, set constant. In this context, a simple consideration
of \secref{sec:intro} led us to the question of finding a lower
bound for the \mw mass, which would be consistent with the
observed radial motions of distant tracers. That question
motivated the main part of our work
(\secrefs{sec:ensemble}{sec:results}), which was aimed at testing
our expectation that the estimated mass value could be reduced
with more general solutions to Jeans equations. We illustrated
this possibility on the example of point mass field, which is
tractable in an analytical way and should approximate to some
extent the galactic gravitational field at large radii. In future
work, we expect an analogous reduction in mass estimates to occur
for more realistic mass profiles.

The principal quantity derivable from a phase space distribution
function (\pdf) and used to infer the total mass is the
theoretical radial velocity dispersion (\rvd) profile that can be
compared with the observed one. In obtaining a \rvd profile from
observations, one should not include gravitationally unbound
bodies. (In obtaining stationary Jeans equations, one assumes that
boundary surface integrals vanish at infinity, that is, no outflow
of matter is possible.) In this context, we gave arguments in
\secref{sec:TrivialEstimator} that a fast receding spheroidal
dwarf galaxy \leoI (and a few other objects) may not be bound to
the \mw and, for this reason, should be rejected at the stage of
preparing the observed \rvd (with \leoI included, i.e., bound to
the \mw, the whole analysis based on Jeans modeling to infer total
mass would become quite meaningless, even when finally predicting
a low mass, since within $250\kpc$ the total mass can be easily
estimated from the escape argument for \leoI to be at least of
$\approx1.2\times10^{12}\msun$, or from the timing argument for
\leoI to be $2-3\times10^{12}\msun$).

In \secref{sec:rvdprofile}, using tracers collected from recent
literature, we performed a simple Monte-Carlo simulation to obtain
a smooth model curve representing the observed \mw's \rvd profile
within estimated uncertainties. We prepared two such model curves
(see \figref{fig:RVDprofiles}), corresponding to two samples of
tracers, \tII and \tI (see \tabref{tab:I}). The first sample was
obtained assuming that both \leoI and a fast and distant BHB star
\textit{J160826.42+065542.3} are not bound to the \mw. The second
sample discards an additional four tracers (including
\textit{Hercules}). We stress, however, that on comparing the two
\rvd profiles (see, \figref{fig:RVDBackgr}), one must arrive at
the conclusion that the inclusion or exclusion of the four tracers
does not influence the \pdf profile significantly, and that only
the BHB star and \textit{Hercules} make a difference. Thus the
exclusion effectively concerned two tracers that were decisive for
the shape of \rvd profile. Either way, the present paper aimed to
show the possibility of reducing mass estimates for a given sample
of tracers. We consider this possibility more important than the
very different problem of which of the three objects (mentioned
above) are gravitationally bound to the \mw.

Applying the point-mass approximation, we hypothesized that most
of the matter in the \mw may be more compactly distributed in
space than thought so far, such that the tracers beyond a boundary
region $20-25\kpc$ could be regarded as test bodies moving in a
nearly point source field. From the standpoint of the distant
tracers, the central mass of the field can be effectively used as
a substitute for true galaxy mass. Accordingly, we appropriately
cut off the support of the \pdf to exclude orbits penetrating the
internal region, where the enclosed galactic mass is not yet
saturated to the value of the substitute central mass, and where
the point-mass approximation may not yet apply. In the boundary
region's vicinity, where the central mass may be higher than the
true mass function, this exclusion enables the theoretical \rvd
profile to be reduced effectively, so that it can overlap with the
\rvd value observed in this region. The mechanism of this
reduction is simple. Elliptical orbits can osculate the boundary
region only with their apocentric parts. As a result, the radial
component of motion is small (with large tangential motion). For
more circular orbits, which mimic the circular motion in the
galactic disk, the radial motion is also minimal. However, the
question of the motion of closer tracers is not that important for
total mass estimations. These tracers carry the information mainly
about the peculiarities in the internal distribution of mass,
rather than on the total mass value. In the internal region (which
we excluded by cutting off the \pdf{}s support) or close to the
boundary region, one could rightly expect some corrections to
occur to the secondary quantities inferred in the point-mass
approximation. But more realistic mass profiles in the interior
cannot change the total mass estimate, which is mainly based on
the radial motions of distant tracers. For a better determination
of the total mass, it is far more important to have a
significantly larger sample of external velocity tracers and
(hopefully in the nearest future) complete measurements of the
transversal motion.

Estimating galactic masses in the point-mass approximation showed
up fruitful and widely used in the literature. Our approach,
however, is essentially different and allows for unrestricted,
mostly general solutions in the point mass field. Instead of
solving Jeans equations involving the secondary quantities related
to some unknown \pdf, we looked for the \pdf directly with the
help of the Keplerian ensemble method we developed in
\secref{sec:ensemble}. Our \pdf satisfies the collision-less
Boltzmann equation and is positive definite by construction. This
way, we avoided the problem characteristic of Jeans modeling that
solutions of Jeans equations do not necessarily lead to physical,
non-negative \pdf{}s. We decided to work in this approximation for
two reasons. Firstly, the spatial orbits are explicitly known, and
some exact formulas in the phase space are obtained more easily
and more straightforwardly than for more complicated mass
profiles. Secondly, the approximation is natural for testing our
hypothesis of compact mass distribution in the \mw.

The Keplerian ensemble method introduced in \secref{sec:ensemble}
shows how to deal with a continuous collection of confocal
elliptical orbits under spherical symmetry and how to use it in
practice. We allowed for a general \pdf being  a function of two
constants of motion and with appropriately cutoff support. The
secondary quantities such as the theoretical \rvd, are obtained
directly from the \pdf and can be quite general functions of the
radial variable. In this method, a particular \pdf is found by
minimizing the discrepancy between theoretical and measured \rvd
profiles. We applied our method to estimating the lower bound for
the \mw mass. To this, we used sample \tI. Based on the
observation in caption to \figref{fig:MassEstimate} in
\secref{sec:TrivialEstimator}, we can expect that  a similar study
for sample \tII would lead to masses that are higher by a factor
of $1.16\approx 1.2$.

Both on the theoretical grounds in \secref{sec:intro} and by
interpreting our numerical results of \secref{sec:results}, we
came to the conclusion that there is no natural upper limit for
the mass estimate: with the same \rvd profile, the \mw mass could
equally well be as high as $10^{12}\msun$ or even higher.  The
absence of such a limit shows that Jeans modeling as a means of
deducing the total gravitating mass from the radial motions alone
is highly underdetermined. (The known mass-anisotropy degeneracy
is a particular example.) As a consequence, the mass is highly
model-dependent and not entirely constrained by measurements of
the radial motions. The criterion for mass must be different.
Based on these observations in the point mass field, one should
expect an even higher degree of arbitrariness to occur in more
general situations of spatially extended mass distributions. The
indeterminacy shown has also an epistemological significance since
the total mass, being a quantity asymptotic by its nature, should
be determinable from motions of external tracers, whereas the only
measurements currently available at large radii (the radial
motions) are scant so cannot be conclusive.

By applying the Keplerian ensemble method, we found in
\secref{sec:results} using sample \tI that a \mw mass as small as
$2.4\times10^{11}\msun$ can still be consistent with the radial
motions of tracers. We chose this example value because it is
greater than the approximate lower bound of about
$2.0\times10^{11}$ determined for sample \tI in
\figref{fig:ChiMass}, and it agrees with the mass value obtained
based on the \mw rotation curve inside $20\kpc$
\citep{1992AJ....103.1552M}. Interestingly enough, the same value
coincides with the total \mw mass obtained by
\citet{1987ApJ...320..493L} in the point mass field for a sample
of distant satellites. Thus, by choosing the \mw mass of
$2.4\times10^{11}\msun$, we would have both consistency with the
\mw's rotation curve out to $20\kpc$ and excellent approximation
of the galactic potential by a point mass field in the region
outside $20\kpc$, where we applied our Keplerian ensemble method
to infer the \mw mass from the radial motion of tracers. We also
presented our prediction for the several secondary quantities
implied by the \pdf found, corresponding to the \rvd at
$M=2.4\times10^{11}\msun$. These observables could be tested when
more accurate data are available at larger radii. In particular,
good measurements of the tracer density profiles would provide an
additional relation to be satisfied by the expansion parameters of
the \pdf in \eqref{eq:expansion}. This would in turn, be helpful
in constraining the allowed bound for the \mw mass in the
point-mass approximation.

The possible low value of the lower bound for the \mw mass is
consistent with measurements in the \mw interior. Dark matter is
insignificant out to $5\kpc$ \citep{2002MNRAS.330..591B}.
Moreover, the dynamical mass implied by the rotation of the \mw
inside solar radius is compatible with the mass in compact objects
ascertained through microlensing measurements
\citep{2012A&A...546A.126S}. But there is also a recent
controversy about whether the mass of \mw's interior could be that
low. Depending on model assumptions, the same data on local
motions of a class of tracers lead to estimates of non-baryonic
dark matter in the solar vicinity differing by a large factor: see
\citet{2012ApJ...751...30M} and the critique in
\citet{2012ApJ...756...89B}. To verify whether the total mass
could indeed be that low requires further studies. As for now,
 one can state with certainty only the main conclusion of this
paper that in the framework of Newtonian mechanics in the point
mass approximation, one can find distribution functions in the
phase space that account for the radial motions of Galaxy
satellites, with much lower mass than thought so far. To see the
extent to which this reduction could be general, it is necessary
to consider more complicated potentials in the future than the
point mass that could be analyzed in a similar manner, assuming
most general phase space without a priori constraints, as this
would provide better estimates for the \mw mass. (As we noted, by
imposing some constraints on the form of solutions to Jeans
equation, one can increase the mass estimate.) We expect that a
reduction in the Galaxy mass estimate similar to the one presented
here for the point mass should also be possible for models
assuming an extended non-baryonic dark halo.

With regard to the model of the phase space developed here, the
next step is to use base functions on the standardized simplex of
the form $\sqrt{w}\,\mathcal{Q}$, with $\mathcal{Q}$ being
polynomials in $\xi$ and $\eta$, and $w={\xi\eta\br{1-\xi-\eta}}$
as the weight in the scalar product. This would ensure that the
phase space distribution function by construction vanishes
smoothly on the boundary of the integration domain in the $\mu_u$
integral in Eq.\eqref{eq:muIntegral}, reducing the number of
spurious circular orbits at outermost radii. It is also desirable
to find best parameters $u_a$ and $u_b$, which are crucial for
defining the phase space. Both these tasks are computationally
more demanding, but they would improve the results. In addition,
to gain more control over secondary quantities in the minimization
procedure, one can try to impose some constraints on these
quantities, provided appropriate measurements are available.

\section*{Acknowledgements}
We would like to thank the anonymous referee for a careful reading
of our manuscript and for many detailed and constructive
suggestions that improved the presentation of this paper. We also
thank James Dwyer for interesting and stimulating discussions.

\bibliography{globular}

\begin{thebibliography}{}

\bibitem[\protect\citeauthoryear{{An} \& {Evans}}{{An} \&
  {Evans}}{2009}]{2009ApJ...701.1500A}
{An} J.~H.,  {Evans} N.~W.,  2009, \apj, 701, 1500

\bibitem[\protect\citeauthoryear{{Bahcall} \& {Tremaine}}{{Bahcall} \&
  {Tremaine}}{1981}]{1981ApJ...244..805B}
{Bahcall} J.~N.,  {Tremaine} S.,  1981, \apj, 244, 805

\bibitem[\protect\citeauthoryear{{Beers}, {Preston} \& {Shectman}}{{Beers}
  et~al.}{1992}]{1992AJ....103.1987B}
{Beers} T.~C.,  {Preston} G.~W.,    {Shectman} S.~A.,  1992, \aj, 103, 1987

\bibitem[\protect\citeauthoryear{{Belokurov}, {Walker}, {Evans}, {Faria},
  {Gilmore}, {Irwin}, {Koposov}, {Mateo}, {Olszewski} \& {Zucker}}{{Belokurov}
  et~al.}{2008}]{2008ApJ...686L..83B}
{Belokurov} V.,  {Walker} M.~G.,  {Evans} N.~W.,  {Faria} D.~C.,  {Gilmore} G.,
   {Irwin} M.~J.,  {Koposov} S.,  {Mateo} M.,  {Olszewski} E.,    {Zucker}
  D.~B.,  2008, \apjl, 686, L83

\bibitem[\protect\citeauthoryear{{Belokurov}, {Walker}, {Evans}, {Gilmore},
  {Irwin}, {Mateo}, {Mayer}, {Olszewski}, {Bechtold} \&
  {Pickering}}{{Belokurov} et~al.}{2009}]{2009MNRAS.397.1748B}
{Belokurov} V.,  {Walker} M.~G.,  {Evans} N.~W.,  {Gilmore} G.,  {Irwin} M.~J.,
   {Mateo} M.,  {Mayer} L.,  {Olszewski} E.,  {Bechtold} J.,    {Pickering} T.,
   2009, \mnras, 397, 1748

\bibitem[\protect\citeauthoryear{{Bissantz} \& {Gerhard}}{{Bissantz} \&
  {Gerhard}}{2002}]{2002MNRAS.330..591B}
{Bissantz} N.,  {Gerhard} O.,  2002, \mnras, 330, 591

\bibitem[\protect\citeauthoryear{{Bovy}, {Hogg} \& {Rix}}{{Bovy}
  et~al.}{2009}]{2009ApJ...704.1704B}
{Bovy} J.,  {Hogg} D.~W.,    {Rix} H.-W.,  2009, \apj, 704, 1704

\bibitem[\protect\citeauthoryear{{Bovy}, {Murray} \& {Hogg}}{{Bovy}
  et~al.}{2010}]{2010ApJ...711.1157B}
{Bovy} J.,  {Murray} I.,    {Hogg} D.~W.,  2010, \apj, 711, 1157

\bibitem[\protect\citeauthoryear{{Bovy} \& {Tremaine}}{{Bovy} \&
  {Tremaine}}{2012}]{2012ApJ...756...89B}
{Bovy} J.,  {Tremaine} S.,  2012, \apj, 756, 89

\bibitem[\protect\citeauthoryear{{Brown}, {Geller}, {Kenyon} \&
  {Diaferio}}{{Brown} et~al.}{2010}]{2010AJ....139...59B}
{Brown} W.~R.,  {Geller} M.~J.,  {Kenyon} S.~J.,    {Diaferio} A.,  2010, \aj,
  139, 59

\bibitem[\protect\citeauthoryear{{Byrd}, {Valtonen}, {McCall} \&
  {Innanen}}{{Byrd} et~al.}{1994}]{1994AJ....107.2055B}
{Byrd} G.,  {Valtonen} M.,  {McCall} M.,    {Innanen} K.,  1994, \aj, 107, 2055

\bibitem[\protect\citeauthoryear{{Carney}}{{Carney}}{1984}]{1984PASP...96..841C}
{Carney} B.~W.,  1984, \pasp, 96, 841

\bibitem[\protect\citeauthoryear{{Clewley}, {Warren}, {Hewett}, {Norris} \&
  {Evans}}{{Clewley} et~al.}{2004}]{2004MNRAS.352..285C}
{Clewley} L.,  {Warren} S.~J.,  {Hewett} P.~C.,  {Norris} J.~E.,    {Evans}
  N.~W.,  2004, \mnras, 352, 285

\bibitem[\protect\citeauthoryear{{Courant} \& {Hilbert}}{{Courant} \&
  {Hilbert}}{1953}]{1953mmp..book.....C}
{Courant} R.,  {Hilbert} D.,  1953, {Methods of mathematical physics - Vol.1;
  Vol.2}

\bibitem[\protect\citeauthoryear{{Deason}, {Belokurov}, {Evans} \&
  {An}}{{Deason} et~al.}{2012}]{2012MNRAS.424L..44D}
{Deason} A.~J.,  {Belokurov} V.,  {Evans} N.~W.,    {An} J.,  2012, \mnras,
  424, L44

\bibitem[\protect\citeauthoryear{{Deason}, {Belokurov}, {Evans}, {Koposov},
  {Cooke}, {Pe{\~n}arrubia}, {Laporte}, {Fellhauer}, {Walker} \&
  {Olszewski}}{{Deason} et~al.}{2012}]{2012MNRAS.425.2840D}
{Deason} A.~J.,  {Belokurov} V.,  {Evans} N.~W.,  {Koposov} S.~E.,  {Cooke}
  R.~J.,  {Pe{\~n}arrubia} J.,  {Laporte} C.~F.~P.,  {Fellhauer} M.,  {Walker}
  M.~G.,    {Olszewski} E.~W.,  2012, \mnras, 425, 2840

\bibitem[\protect\citeauthoryear{{Deason}, {McCarthy}, {Font}, {Evans},
  {Frenk}, {Belokurov}, {Libeskind}, {Crain} \& {Theuns}}{{Deason}
  et~al.}{2011}]{2011MNRAS.415.2607D}
{Deason} A.~J.,  {McCarthy} I.~G.,  {Font} A.~S.,  {Evans} N.~W.,  {Frenk}
  C.~S.,  {Belokurov} V.,  {Libeskind} N.~I.,  {Crain} R.~A.,    {Theuns} T.,
  2011, \mnras, 415, 2607

\bibitem[\protect\citeauthoryear{{Dehnen} \& {Binney}}{{Dehnen} \&
  {Binney}}{1998}]{1998MNRAS.294..429D}
{Dehnen} W.,  {Binney} J.,  1998, \mnras, 294, 429

\bibitem[\protect\citeauthoryear{{Di Cintio}, {Knebe}, {Libeskind}, {Brook},
  {Yepes}, {Gottl{\"o}ber} \& {Hoffman}}{{Di Cintio}
  et~al.}{2013}]{2013MNRAS.431.1220D}
{Di Cintio} A.,  {Knebe} A.,  {Libeskind} N.~I.,  {Brook} C.,  {Yepes} G.,
  {Gottl{\"o}ber} S.,    {Hoffman} Y.,  2013, \mnras, 431, 1220

\bibitem[\protect\citeauthoryear{{Dohm-Palmer}, {Helmi}, {Morrison}, {Mateo},
  {Olszewski}, {Harding}, {Freeman}, {Norris} \& {Shectman}}{{Dohm-Palmer}
  et~al.}{2001}]{2001ApJ...555L..37D}
{Dohm-Palmer} R.~C.,  {Helmi} A.,  {Morrison} H.,  {Mateo} M.,  {Olszewski}
  E.~W.,  {Harding} P.,  {Freeman} K.~C.,  {Norris} J.,    {Shectman} S.~A.,
  2001, \apjl, 555, L37

\bibitem[\protect\citeauthoryear{{Francis} \& {Anderson}}{{Francis} \&
  {Anderson}}{2009}]{2009NewA...14..615F}
{Francis} C.,  {Anderson} E.,  2009, \na, 14, 615

\bibitem[\protect\citeauthoryear{{Geha}, {Willman}, {Simon}, {Strigari},
  {Kirby}, {Law} \& {Strader}}{{Geha} et~al.}{2009}]{2009ApJ...692.1464G}
{Geha} M.,  {Willman} B.,  {Simon} J.~D.,  {Strigari} L.~E.,  {Kirby} E.~N.,
  {Law} D.~R.,    {Strader} J.,  2009, \apj, 692, 1464

\bibitem[\protect\citeauthoryear{{Harris}}{{Harris}}{1996}]{1996AJ....112.1487H}
{Harris} W.~E.,  1996, \aj, 112, 1487

\bibitem[\protect\citeauthoryear{{Jeans}}{{Jeans}}{1915}]{1915MNRAS..76...70J}
{Jeans} J.~H.,  1915, \mnras, 76, 70

\bibitem[\protect\citeauthoryear{{Kahn} \& {Woltjer}}{{Kahn} \&
  {Woltjer}}{1959}]{1959ApJ...130..705K}
{Kahn} F.~D.,  {Woltjer} L.,  1959, \apj, 130, 705

\bibitem[\protect\citeauthoryear{{Klypin}, {Zhao} \& {Somerville}}{{Klypin}
  et~al.}{2002}]{2002ApJ...573..597K}
{Klypin} A.,  {Zhao} H.,    {Somerville} R.~S.,  2002, \apj, 573, 597

\bibitem[\protect\citeauthoryear{{Koch}, {Wilkinson}, {Kleyna}, {Irwin},
  {Zucker}, {Belokurov}, {Gilmore}, {Fellhauer} \& {Evans}}{{Koch}
  et~al.}{2009}]{2009ApJ...690..453K}
{Koch} A.,  {Wilkinson} M.~I.,  {Kleyna} J.~T.,  {Irwin} M.,  {Zucker} D.~B.,
  {Belokurov} V.,  {Gilmore} G.~F.,  {Fellhauer} M.,    {Evans} N.~W.,  2009,
  \apj, 690, 453

\bibitem[\protect\citeauthoryear{{Koposov}, {Rix} \& {Hogg}}{{Koposov}
  et~al.}{2010}]{2010ApJ...712..260K}
{Koposov} S.~E.,  {Rix} H.-W.,    {Hogg} D.~W.,  2010, \apj, 712, 260

\bibitem[\protect\citeauthoryear{{Little} \& {Tremaine}}{{Little} \&
  {Tremaine}}{1987}]{1987ApJ...320..493L}
{Little} B.,  {Tremaine} S.,  1987, \apj, 320, 493

\bibitem[\protect\citeauthoryear{{LUX Collaboration}}{{LUX
  Collaboration}}{2013}]{2013arXiv1310.8214L}
{LUX Collaboration} 2013, ArXiv e-prints 1310.8214

\bibitem[\protect\citeauthoryear{{Magorrian}}{{Magorrian}}{2013}]{2013arXiv1303.6099M}
{Magorrian} J.,  2013, ArXiv e-prints 1303.6099

\bibitem[\protect\citeauthoryear{{Martin}, {Ibata}, {Chapman}, {Irwin} \&
  {Lewis}}{{Martin} et~al.}{2007}]{2007MNRAS.380..281M}
{Martin} N.~F.,  {Ibata} R.~A.,  {Chapman} S.~C.,  {Irwin} M.,    {Lewis}
  G.~F.,  2007, \mnras, 380, 281

\bibitem[\protect\citeauthoryear{{Mateo}}{{Mateo}}{1998}]{1998ARA&A..36..435M}
{Mateo} M.~L.,  1998, \araa, 36, 435

\bibitem[\protect\citeauthoryear{{Merrifield}}{{Merrifield}}{1992}]{1992AJ....103.1552M}
{Merrifield} M.~R.,  1992, \aj, 103, 1552

\bibitem[\protect\citeauthoryear{{Moni Bidin}, {Carraro}, {M{\'e}ndez} \&
  {Smith}}{{Moni Bidin} et~al.}{2012}]{2012ApJ...751...30M}
{Moni Bidin} C.,  {Carraro} G.,  {M{\'e}ndez} R.~A.,    {Smith} R.,  2012,
  \apj, 751, 30

\bibitem[\protect\citeauthoryear{{Morrison}, {Mateo}, {Olszewski}, {Harding},
  {Dohm-Palmer}, {Freeman}, {Norris} \& {Morita}}{{Morrison}
  et~al.}{2000}]{2000AJ....119.2254M}
{Morrison} H.~L.,  {Mateo} M.,  {Olszewski} E.~W.,  {Harding} P.,
  {Dohm-Palmer} R.~C.,  {Freeman} K.~C.,  {Norris} J.~E.,    {Morita} M.,
  2000, \aj, 119, 2254

\bibitem[\protect\citeauthoryear{{Perryman}, {de Boer}, {Gilmore}, {H{\o}g},
  {Lattanzi}, {Lindegren}, {Luri}, {Mignard}, {Pace} \& {de Zeeuw}}{{Perryman}
  et~al.}{2001}]{2001A&A...369..339P}
{Perryman} M.~A.~C.,  {de Boer} K.~S.,  {Gilmore} G.,  {H{\o}g} E.,  {Lattanzi}
  M.~G.,  {Lindegren} L.,  {Luri} X.,  {Mignard} F.,  {Pace} O.,    {de Zeeuw}
  P.~T.,  2001, \aap, 369, 339

\bibitem[\protect\citeauthoryear{{Sakamoto}, {Chiba} \& {Beers}}{{Sakamoto}
  et~al.}{2003}]{2003A&A...397..899S}
{Sakamoto} T.,  {Chiba} M.,    {Beers} T.~C.,  2003, \aap, 397, 899

\bibitem[\protect\citeauthoryear{{Sales}, {Navarro}, {Abadi} \&
  {Steinmetz}}{{Sales} et~al.}{2007}]{2007MNRAS.379.1475S}
{Sales} L.~V.,  {Navarro} J.~F.,  {Abadi} M.~G.,    {Steinmetz} M.,  2007,
  \mnras, 379, 1475

\bibitem[\protect\citeauthoryear{{Sch{\"o}nrich}, {Binney} \&
  {Dehnen}}{{Sch{\"o}nrich} et~al.}{2010}]{2010MNRAS.403.1829S}
{Sch{\"o}nrich} R.,  {Binney} J.,    {Dehnen} W.,  2010, \mnras, 403, 1829

\bibitem[\protect\citeauthoryear{{Sikora}, {Bratek}, {Ja{\l}ocha} \&
  {Kutschera}}{{Sikora} et~al.}{2012}]{2012A&A...546A.126S}
{Sikora} S.,  {Bratek} {\L}.,  {Ja{\l}ocha} J.,    {Kutschera} M.,  2012, \aap,
  546, A126

\bibitem[\protect\citeauthoryear{{Simon} \& {Geha}}{{Simon} \&
  {Geha}}{2007}]{2007ApJ...670..313S}
{Simon} J.~D.,  {Geha} M.,  2007, \apj, 670, 313

\bibitem[\protect\citeauthoryear{{Sohn}, {Besla}, {van der Marel},
  {Boylan-Kolchin}, {Majewski} \& {Bullock}}{{Sohn}
  et~al.}{2013}]{2013ApJ...768..139S}
{Sohn} S.~T.,  {Besla} G.,  {van der Marel} R.~P.,  {Boylan-Kolchin} M.,
  {Majewski} S.~R.,    {Bullock} J.~S.,  2013, \apj, 768, 139

\bibitem[\protect\citeauthoryear{{Starkenburg}, {Helmi}, {Morrison}, {Harding},
  {van Woerden}, {Mateo}, {Olszewski}, {Sivarani}, {Norris}, {Freeman},
  {Shectman}, {Dohm-Palmer}, {Frey} \& {Oravetz}}{{Starkenburg}
  et~al.}{2009}]{2009ApJ...698..567S}
{Starkenburg} E.,  {Helmi} A.,  {Morrison} H.~L.,  {Harding} P.,  {van Woerden}
  H.,  {Mateo} M.,  {Olszewski} E.~W.,  {Sivarani} T.,  {Norris} J.~E.,
  {Freeman} K.~C.,  {Shectman} S.~A.,  {Dohm-Palmer} R.~C.,  {Frey} L.,
  {Oravetz} D.,  2009, \apj, 698, 567

\bibitem[\protect\citeauthoryear{{van der Marel} \& {Guhathakurta}}{{van der
  Marel} \& {Guhathakurta}}{2008}]{2008ApJ...678..187V}
{van der Marel} R.~P.,  {Guhathakurta} P.,  2008, \apj, 678, 187

\bibitem[\protect\citeauthoryear{{Vera-Ciro}, {Helmi}, {Starkenburg} \&
  {Breddels}}{{Vera-Ciro} et~al.}{2013}]{2013MNRAS.428.1696V}
{Vera-Ciro} C.~A.,  {Helmi} A.,  {Starkenburg} E.,    {Breddels} M.~A.,  2013,
  \mnras, 428, 1696

\bibitem[\protect\citeauthoryear{{Wang}, {Frenk}, {Navarro}, {Gao} \&
  {Sawala}}{{Wang} et~al.}{2012}]{2012MNRAS.424.2715W}
{Wang} J.,  {Frenk} C.~S.,  {Navarro} J.~F.,  {Gao} L.,    {Sawala} T.,  2012,
  \mnras, 424, 2715

\bibitem[\protect\citeauthoryear{{Watkins}, {Evans} \& {An}}{{Watkins}
  et~al.}{2010}]{2010MNRAS.406..264W}
{Watkins} L.~L.,  {Evans} N.~W.,    {An} J.~H.,  2010, \mnras, 406, 264

\bibitem[\protect\citeauthoryear{{Wilhelm}, {Beers}, {Sommer-Larsen}, {Pier},
  {Layden}, {Flynn}, {Rossi} \& {Christensen}}{{Wilhelm}
  et~al.}{1999}]{1999AJ....117.2329W}
{Wilhelm} R.,  {Beers} T.~C.,  {Sommer-Larsen} J.,  {Pier} J.~R.,  {Layden}
  A.~C.,  {Flynn} C.,  {Rossi} S.,    {Christensen} P.~R.,  1999, \aj, 117,
  2329

\bibitem[\protect\citeauthoryear{{Wilkinson} \& {Evans}}{{Wilkinson} \&
  {Evans}}{1999}]{1999MNRAS.310..645W}
{Wilkinson} M.~I.,  {Evans} N.~W.,  1999, \mnras, 310, 645

\bibitem[\protect\citeauthoryear{{Xue}, {Rix}, {Zhao}, {Re Fiorentin}, {Naab},
  {Steinmetz}, {van den Bosch}, {Beers}, {Lee}, {Bell}, {Rockosi}, {Yanny},
  {Newberg}, {Wilhelm}, {Kang}, {Smith} \& {Schneider}}{{Xue}
  et~al.}{2008}]{2008ApJ...684.1143X}
{Xue} X.~X.,  {Rix} H.~W.,  {Zhao} G.,  {Re Fiorentin} P.,  {Naab} T.,
  {Steinmetz} M.,  {van den Bosch} F.~C.,  {Beers} T.~C.,  {Lee} Y.~S.,  {Bell}
  E.~F.,  {Rockosi} C.,  {Yanny} B.,  {Newberg} H.,  {Wilhelm} R.,  {Kang} X.,
  {Smith} M.~C.,    {Schneider} D.~P.,  2008, \apj, 684, 1143

\bibitem[\protect\citeauthoryear{{Zaritsky}, {Olszewski}, {Schommer},
  {Peterson} \& {Aaronson}}{{Zaritsky} et~al.}{1989}]{1989ApJ...345..759Z}
{Zaritsky} D.,  {Olszewski} E.~W.,  {Schommer} R.~A.,  {Peterson} R.~C.,
  {Aaronson} M.,  1989, \apj, 345, 759

\end{thebibliography}
\bibliographystyle{mn2e}
\end{document}